\documentclass[prx,floatfix,superscriptaddress,longbibliography,noshowpacs,twocolumn]{revtex4}
\usepackage{graphics,lscape,graphicx,epsfig,amsmath,amssymb,epstopdf,wasysym,comment}
\usepackage[T1]{fontenc}
\usepackage{frcursive}
\usepackage{color}
\newenvironment{frcseries}{\fontfamily{frc}\selectfont}{}

\usepackage{subfigure}
\allowdisplaybreaks

\newcommand{\be}{\begin{equation}}
\newcommand{\ee}{\end{equation}}
\newcommand{\bea}{\begin{eqnarray}}
\newcommand{\eea}{\end{eqnarray}}
\newcommand{\ba}{\begin{eqnarray*}}
\newcommand{\ea}{\end{eqnarray*}}
\newcommand{\dagga}{{\phantom{\dagger}}}
\newcommand{\bR}{\mathbf{R}}

\newcommand{\bK}{\mathbf{K}}
\newcommand{\bq}{\mathbf{q}}

\newcommand{\bk}{\mathbf{k}}

\newcommand{\bp}{\mathbf{p}}

\newcommand{\br}{\mathbf{r}}

\newcommand{\dis}{\displaystyle}

\newcommand{\fract}[2]{\frac{\dis #1}{\dis #2}}

\newcommand{\eqn}[1]{(\ref{#1})}

\newcommand{\bba}{\mathbf{a}}
\newcommand{\bbb}{\mathbf{b}}

\newcommand{\bGamma}{{\boldsymbol{\Gamma}}}
\newcommand{\Cg}[1]{\text{C}_{#1}}
\newcommand{\mC}[1]{\mathcal{C}_{#1}}
\newenvironment{eqs}%
{\begin{equation} \begin{aligned}}%
{\end{aligned} \end{equation} }
\newcommand{\beal}{\begin{eqs}}
\newcommand{\eal}{\end{eqs}}
\newcommand{\bw}{\begin{widetext}}
\newcommand{\ew}{\end{widetext}}
\newcommand{\esp}[1]{\text{e}^{#1}}

\newcommand{\bM}{\mathbf{M}}

\begin{document}

\title{Valley Jahn-Teller effect in twisted bilayer graphene}

\author{M.~Angeli}
\affiliation{International School for
  Advanced Studies (SISSA), Via Bonomea
  265, I-34136 Trieste, Italy} 
  
 \author{E.~Tosatti}
\affiliation{International School for
  Advanced Studies (SISSA), Via Bonomea
  265, I-34136 Trieste, Italy} 
\affiliation{CNR-IOM Democritos, Istituto Officina dei Material, Consiglio Nazionale delle Ricerche} 
\affiliation{International Centre for Theoretical Physics (ICTP), Strada Costiera 11, I-34151 Trieste, Italy}

\author{M.~Fabrizio}
\affiliation{International School for
  Advanced Studies (SISSA), Via Bonomea
  265, I-34136 Trieste, Italy} 

\begin{abstract}
The surprising insulating and superconducting states of narrow-band graphene twisted bilayers have been mostly discussed so far in terms of strong electron correlation, with little or no attention to phonons and electron-phonon effects. We found that, among the 33492 phonons of a fully relaxed $\theta=1.08^\circ$ twisted bilayer, there are few special, hard and nearly dispersionless modes that resemble global vibrations of the moir\'e supercell,  as if it were a single, ultralarge  molecule. One of them, doubly degenerate at $\bGamma$ with symmetry $A_1+B_1$,  couples very strongly with the valley degrees of freedom, also doubly degenerate, realizing a so-called $\text{E}\otimes\text{e}$ Jahn-Teller (JT) coupling.  The JT  coupling lifts very  efficiently all degeneracies which arise from the valley symmetry, and may lead, for an average atomic displacement as small as $0.5~\text{m\AA}$, to an insulating state at charge neutrality. This insulator 
possesses a non-trivial topology testified by the odd winding of the Wilson loop. In addition, freezing the same phonon at a zone boundary point brings about insulating states at most integer occupancies of the four ultra-flat electronic bands.   Following that line, we  further study the properties of the superconducting state that might be stabilized by these modes. Since the JT coupling modulates the hopping between AB and BA stacked regions, pairing occurs in the spin-singlet Cooper channel at the inter-(AB-BA) scale,  which may  condense a superconducting order parameter in the extended $s$-wave and/or $d\pm id$-wave symmetry.

\end{abstract}

\date{April 12, 2019}          
                         
\maketitle

\section{Introduction}

The recent discovery of superconductivity in \textit{magic angle}
($\theta \approx 1.1^\circ$) twisted bilayer graphene (tBLG) 
\cite{Herrero-1,Herrero-2,Yankowitz,Efetov} has stimulated 
an intense theoretical and experimental research activity.  This unexpected phenomenon occurs upon slightly doping insulating states found at fractional fillings, 
the latter
contradicting the metallic behaviour predicted by band structure calculations. Despite the huge size of the unit cell, containing more than $\approx 11000$ atoms, the band structure of tBLG at the first magic angle has been computed with a variety of methods, including tight-binding \cite{Trambly,Trambly2,Shallcross,Sboychakov,Morell,Fabrizio}, continuum models \cite{MacDonald-PNAS2011,Vishvanath-continuum} and DFT \cite{Bernevig_topo,Procolo}.
These approaches predict that all the exotic properties mentioned above arise from four extremely flat bands (FBs), located around the charge neutrality point, 
with a bandwidth of the order of $\approx 10-20$ meV. Owing to the flatness of these bands, the fractional filling insulators found in \cite{Herrero-1,Herrero-2} are \textit{conjectured} to be Mott Insulators,
even though rather anomalous 
ones, 
since
they turn frankly metallic above a critical temperature or above a threshold Zeeman splitting in a magnetic field, features not expected from a Mott insulator. 
Actually, the linear size of the unit cell at the magic angle is as large as $\approx 14$ nm, and the effective \textit{on-site} Coulomb repulsion, the so-called Hubbard $U$, must be given
by the charging energy in this large supercell projected onto the FBs, including screening effects due to the gates and to the other bands. Even 
neglecting the latter, the estimated $U\sim 9~\text{meV}$ is comparable to the bandwidth of the FBs \cite{Vafek-2}. Since the FBs are reproducibly 
found in experiments \cite{Efetov,Herrero-1,Herrero-2,Yankowitz,Pasupathy,Perge,Polshyn,Herrero-3,Ashoori} to be separated from the other bands by a 
gap of around $\sim 30-50$ meV, the actual value of $U$ should be significantly smaller, implying that tBLG might
not be more correlated than a single graphene sheet \cite{U-graphene}. In turn, this suggests that the insulating behaviour at $\nu=\pm2$ occupancy might 
instead
be the result of a weak-coupling Stoner 
or CDW band instability driven by electron-electron and/or electron-phonon interactions, rather than a Mott localization phenomenon.\\
As pointed out in \cite{Senthil-2}, in order to open an insulating gap the band instability must break the twofold degeneracy at the $\bK$-points imposed by the $D_6$ space group symmetry of the 
moir\'e superlattice, as well as the additional twofold degeneracy due to the so-called valley charge conservation. This conserved quantity is associated with an emergent dynamical $U(1)$ symmetry that appears at small twist angles,
a symmetry 
which, unlike spatial symmetries, is rather subtle and elusive. It is therefore essential to identify a microscopic mechanism that could efficiently break this emergent symmetry, hereafter refereed to as $U_v(1)$.  The most natural candidate is 
the Coulomb repulsion \cite{Senthil-2,McDonald_HF,Vishvanath_SC_IVC}, whose Fourier transform decays more slowly than that of electron hopping, possibly introducing a non negligible coupling among the two valleys even at small twist angles. Indeed DFT-based calculations show 
a tiny valley splitting \cite{Bernevig_topo,Procolo}, almost at the limits of accuracy of the method, which is nevertheless too small to explain the insulating states found in the FBs.\\
Here we uncover another $U_v(1)$-breaking mechanism involving instead the lattice degrees of freedom, mostly ignored so far. 
It must be recalled
 that ab-initio DFT-based calculations fail to predict well defined
FBs separated from other bands unless atomic positions are allowed to relax \cite{Nam_Koshino_PRB,Kaxiras-1,Fabrizio,Procolo,Kaxiras-2,Choi}, especially out-of-plane. That
alone
already demonstrated that the effects of atomic motions in the lattice
are not at all negligible in tBLG,  
further supported
by the significant phonon contribution to transport \cite{Polshyn,Sarma,Vignale}. 
We calculate the phonon spectrum of the fully-relaxed bilayer at $1.08^\circ$ twist angle, which shows the presence, among the about thirty thousands phonons, of a small set of very special optical modes, with C-C stretching character, very narrow and 
uniquely
coherent over the moir\'e supercell Brillouin Zone. Among them, we find a doubly degenerate optical mode that couples to the $U_v(1)$ symmetry 
much
more efficiently than 
Coulomb repulsion 
seems to do
 in DFT calculations.
A subsequent 
frozen-phonon tight-binding calculation shows that this mode is able to fully lift the valley degeneracy even when its lattice deformation amplitude is extremely small. Remarkably, 
both electrons and phonons are twofold  $U_v(1)$ -degenerate, and 
the coupling of this mode with the electron bands actually realizes an $\text{E}\otimes\text{e}$ Jahn-Teller (JT) effect \cite{Englman}.
This effect is able to stabilize insulating states at integer occupancies of the FBs, both even and odd.
Moreover, a surprising and important additional result will be that the electron-phonon coupling magnitude controlling this process is extremely large, and not small as one could generally expect for a very narrow band. 
We 
conclude
by studying the superconducting state that might be mediated by the Jahn-Teller coupling in a minimal tight-binding model of the FBs that reproduces symmetries and topological properties of the realistic band structure calculations.  \\

The work is organized as follows. In section \ref{Section II} we specify the geometry of the tBLG studied and define useful quantities that are used throughout the article. In section 
\ref{Lattice relaxation and band structure} we briefly discuss the band structure obtained by a realistic tight-binding calculation of the tBLG with fully relaxed atomic positions. The phonon spectrum and its properties, especially focusing on special optical modes strongly coupled with the valley $U_v(1)$ symmetry, are throughly discussed in section \ref{Phonons in twisted bilayer graphene}. Section \ref{Phonon mediated superconductivity} addresses the properties of the superconducting state that might be stabilized by the particular phonon mode identified in the previous section, through a mean-field calculation using a model tight-binding 
Hamiltonian of the FBs.  Finally, section \ref{Conclusions} is devoted to concluding remarks.

\begin{figure}
\centerline{\includegraphics[width=0.5\textwidth]{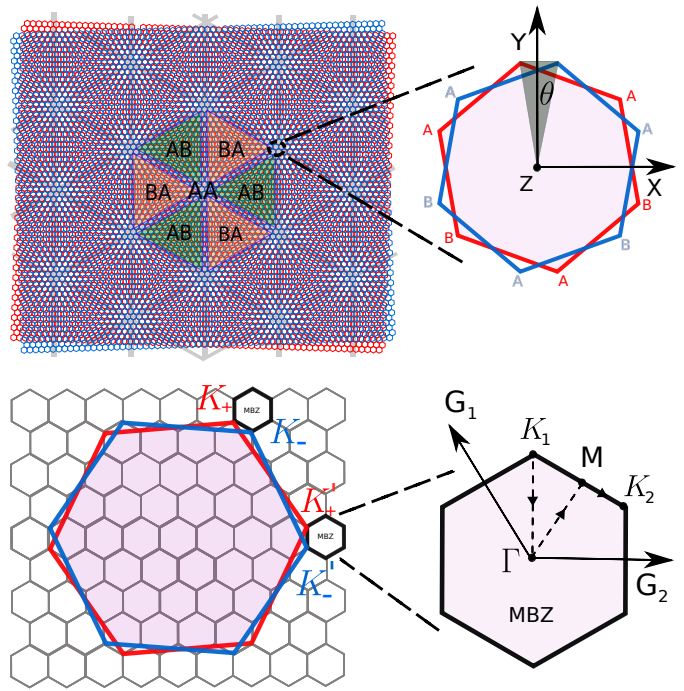}}
\caption{Top panel: moir\'e superlattice formed by two unrelaxed graphene layers (in blue and red) twisted by an angle $\theta$. We indicate the different stacking regions: AA and the two different Bernal regions AB and BA. The domain walls (DWs) separate AB from BA regions and connect different AA regions.  On the right, a zoom into the AA stacked region is shown: two overlapping hexagons in distinct graphene layers are rotated in opposite directions around the perpendicular $z$-axis by an angle $\theta/2$. Bottom panel: the folding procedure in tBLG. The original single layer Brillouin zones (in red and blue) are folded into the mini-Brillouin zone (MBZ). Two inequivalent $\bK$ points in different layers ($\bK_-$ and $\bK_+'$ or $\bK_-'$ and $\bK_+$) are folded into the same points, $\bK_1$ or $\bK_2$, of the MBZ. The path $\bK_1\to\bGamma\to\bM\to \bK_2$ is also shown.}
\label{lattice}
\end{figure}

\section{moir\'e superlattice and symmetries}
\label{Section II}
In Fig.~\ref{lattice} we show the geometry of the tBLG that we shall use hereafter. 
We start from a AA stacked bilayer and rotate in opposite directions the two 
layers around the center of a hexagon by an angle $1.08^\circ/2$, leading to a unit cell with 11164 carbon atoms. The moir\'e superlattice, top left panel, with the reference frame defined in the top right panel, possesses a full $D_6$ spatial symmetry. 
Specifically, resolving the action of each symmetry operation in the indices that identify the two 
layers, 1 and 2, the two sublattices within each layer, A and B, and, finally, the two sublattices AB and BA of the moir\'e superlattice (see bottom panel of Fig.~\ref{lattice}), we have that:
\begin{itemize}
\item the rotation $\Cg{3z}$ by $120^\circ$ degrees around the $z$-axis is diagonal in all indices, 1 and 2, A and B, AB and BA;
\item the $\Cg{2x}$ rotation by 
$180^\circ$ degrees around the $x$-axis interchanges 1 with 2, A with B, but is 
diagonal in AB and BA;
\item the $\Cg{2y}$ rotation by 
$180^\circ$ degrees around the $y$-axis interchanges 1 with 2, AB with BA, but is diagonal in A and B;
\item finally, the action of a $\Cg{2z}$ rotation by 
$180^\circ$ degrees around the $z$-axis is
a composite symmetry operation
obtained by noting that 
$\Cg{2z}=\Cg{2x}\times\Cg{2y}$.
\end{itemize} 
In Table~\ref{irreps} we list the irreducible representations (irreps) of 
the $D_6$ space group and the action on each of them of the symmetry transformations $\Cg{3z}$, $\Cg{2x}$ and $\Cg{2y}$.

\begin{table}[thb]
\vspace{0.2cm}
\begin{tabular}{|c|c|c|c|}\hline
~ & $\Cg{3z}$ & $\Cg{2x}$ & $\Cg{2y}$ \\ \hline
A$_1(1)$   & +1 & +1 & +1  \\ \hline
A$_2(1)$   & +1 & -1  & -1   \\ \hline
B$_1(1)$   & +1  & +1 & -1  \\ \hline
B$_2(1)$   & +1  & -1  & +1  \\ \hline
E$_1(2)$   &
$\begin{pmatrix}
\cos\phi & -\sin\phi\\
\sin\phi & \cos\phi
\end{pmatrix} $ &
$\begin{pmatrix}
+1 & 0\\
0 & -1
\end{pmatrix} $ &
$\begin{pmatrix}
-1 & 0\\
0 & +1
\end{pmatrix} $  \\ \hline
E$_2(2)$ &
$\begin{pmatrix}
\cos\phi & -\sin\phi\\
\sin\phi & \cos\phi
\end{pmatrix}$ &
$\begin{pmatrix}
1 & 0\\
0 & -1
\end{pmatrix} $ &
$\begin{pmatrix}
1 & 0\\
0 & -1
\end{pmatrix} $  \\ \hline
\end{tabular}
\caption{
Nontrivial 
irreducible representations of the space group $D_6$. Each representation has the degeneracy shown in parenthesis. We also list the action of the symmetry operations for each representation, where $\phi=2\pi/3$.
 } 
\label{irreps}
\end{table}  

\subsection{$U_v(1)$ valley symmetry}
\label{$U_v(1)$ valley symmetry}
The mini Brillouin zone (MBZ) that corresponds to the real space geometry of Fig.~\ref{lattice} 
and its relationship with the original graphene Brillouin zones are shown in the bottom panels of that figure (for better readability, at a larger angle than the actual $1.08^\circ$ which we use). Because of the chosen geometry, the Dirac point $K_+$($K'_+$) of the top layer and $K'_-$($K_-$) of the bottom one fold onto the same point, $K_1$ or $K_2$, of the MBZ, so that a finite matrix element of the Hamiltonian between them is allowed by symmetry.  
Nevertheless, as pointed out by \cite{MacDonald-PNAS2011}, the matrix element of the one-body component of the Hamiltonian is negligibly small at small twist angles, so that the two Dirac points, hereafter named \textit{valleys}, remain effectively independent of each other. This implies that the operator 
\beal
\Delta N_v &= N_1 - N_2\,,
\label{tau_3}
\eal
where $N_1$ and $N_2$ are the occupation numbers of each valley, must commute with the non-interacting Hamiltonian of the tBLG at small angles. That operator is in fact the generator of the $U_v(1)$ symmetry.  
As we will see in the following (see \ref{accidental}), the interplay between the valley symmetry and $C_{2y}$ is responsible for the additional two-fold degeneracies beyond those of Table~\ref{irreps} in both electron band structure and phonon spectra along all the points in the MBZ invariant under that symmetry. 

\subsection{Wannier orbitals}
We shall assume in the following Wannier orbitals (WOs) that are centered at the Wyckoff positions $2c$ of the moir\'e triangular superlattice, i.e., at the AB and BA region centers, even though their probability distribution has
a 
very
substantial component in other regions such as AA. The large size of the unit cell has originated so far 
an intense effort to find a minimal model faithfully describing the FBs physics. While a variety of WOs centered at different Wyckoff 
positions has been proposed \cite{Mellado,Senthil-1,Senthil-2,Vafek_PRX,Vishvanath_WO,Fu_PRX,Yuan,Bernevig_topo,Kaxiras-2}, our simple assumption is well suited for our purposes. 
The site symmetry at the Wyckoff positions $2c$ is $D_3$, and includes only $\Cg{3z}$ and $\Cg{2x}$ with irreps $A_1$, $A_2$ and $E$. Inspired by the symmetries of the Bloch states at the high-symmetry points \cite{Fabrizio}, see Sec.~\ref{Lattice relaxation and band structure}), we consider two $A_1$ and $A_2$ one-dimensional irreps (1d-irreps), both invariant under $\Cg{3z}$ 
and eigenstates of  $\Cg{2x}$ with opposite eigenvalues $c_{2x}=\pm 1$. In addition, we consider
one two-dimensional irrep (2d-irrep) $E$, which transforms under $\Cg{3z}$ and $\Cg{2x}$ as the 
2d-irreps in Table~\ref{irreps}, hence comprises eigenstates of $\Cg{2x}$, with 
opposite eigenvalues $c_{2x}=\pm 1$, which are not invariant under $\Cg{3z}$. \\
We define on sublattice $\alpha=AB,BA$ the spin-$\sigma$ WO 
annihilation operators $\Psi_{\alpha,\bR\sigma}$ and $\Phi_{\alpha,\bR\sigma}$ corresponding to the two 1d-irreps ( $\Psi$) or the single 2d-irrep ($\Phi$), respectively
\beal
\Psi^\dagga_{\alpha,\bR\sigma} &= \begin{pmatrix}
\Psi_{\alpha,1,s,\bR\sigma}\\
\Psi_{\alpha,2,s,\bR\sigma}\\
\Psi_{\alpha,1,p,\bR\sigma}\\
\Psi_{\alpha,2,p,\bR\sigma}
\end{pmatrix}\,,&
\Phi^\dagga_{\alpha,\bR\sigma} &= \begin{pmatrix}
\Phi_{\alpha,1,s,\bR\sigma}\\
\Phi_{\alpha,2,s,\bR\sigma}\\
\Phi_{\alpha,1,p,\bR\sigma}\\
\Phi_{\alpha,2,p,\bR\sigma}
\end{pmatrix}\,, \label{spinors}
\eal
where the subscript $s$ refers to $c_{2x}=+1$, and $p$ to $c_{2x}=-1$, while the labels 1 and 2  
refer to the two valleys. It is implicit that each component is itself a spinor that includes fermionic operators corresponding to different WOs that transform like the same irrep. We shall combine the operators of different sublattices into a single spinor 
\beal
\Psi^\dagga_{\bR\sigma} &= \begin{pmatrix}
\Psi_{AB,\bR\sigma}\\
\Psi_{BA,\bR\sigma}
\end{pmatrix}\,,&
\Phi^\dagga_{\bR\sigma} &= \begin{pmatrix}
\Phi_{AB,\bR\sigma}\\
\Phi_{BA,\bR\sigma}
\end{pmatrix}\,.\label{spinors-2}
\eal 
We further introduce three different Pauli matrices $\sigma_a$  
that act in the moir\'e sublattice space (AB, BA), $\mu_a$ in the $c_{2x}=\pm 1$ space ($s$, $p$), and $\tau_a$ in the valley space (1, 2), 
where $a=0,1,2,3$, 
$a=0$ denoting the identity. \\
With these definitions, the generator \eqn{tau_3} of the valley $U_v(1)$ symmetry becomes simply
\beal
\Delta N_v &= \sum_{\bR\sigma}\, \Big(\,\Psi^\dagger_{\bR\sigma}\,\sigma_0\,\tau_3\,\mu_0\,\,\Psi^\dagga_{\bR\sigma}
+ \Phi^\dagger_{\bR\sigma}\,\sigma_0\,\tau_3\,\mu_0\,\Phi^\dagga_{\bR\sigma}\,\Big)\,.
\label{tau-3-bis}
\eal

It is now worth deriving the expression of the space group symmetry operations in this notation and representation.By definition, the $\Cg{2x}$ transformation corresponds to the simple operator 
\beal
\mC{2x}\Big(\Psi^\dagga_{\bR\sigma}\Big) &= \sigma_0\,\tau_0\,\mu_3\, 
\Psi^\dagga_{ \Cg{2x}(\bR)\,\sigma}\,,\\
\mC{2x}\Big(\Phi^\dagga_{\bR\sigma}\Big) &= \sigma_0\,\tau_0\,\mu_3\, 
\Phi^\dagga_{ \Cg{2x}(\bR)\,\sigma}\,.
\eal
The $180^\circ$ rotation around 
the $z$-axis that connects sublattice AB with BA of each layer ($C_{2z}$) is not diagonal in the valley indices and can be represented by  \cite{Bernevig_topo,Senthil-1}
\beal
\mC{2z}\Big(\Psi^\dagga_{\bR\sigma}\Big) &= \sigma_1\,\tau_1\,\mu_0\, 
\Psi^\dagga_{ \Cg{2z}(\bR)\,\sigma}\,,\\
\mC{2z}\Big(\Phi^\dagga_{\bR\sigma}\Big) &= \sigma_1\,\tau_1\,\mu_0\, 
\Phi^\dagga_{ \Cg{2z}(\bR)\,\sigma}\,.
\eal 
Finally, since 
$\Cg{2y}=\Cg{2z}\times\Cg{2x}$, then  
\beal
\mC{2y}\Big(\Psi^\dagga_{\bR\sigma}\Big) &= \sigma_1\,\tau_1\,\mu_3\, 
\Psi^\dagga_{ \Cg{2y}(\bR)\,\sigma}\,,\\
\mC{2y}\Big(\Phi^\dagga_{\bR\sigma}\Big) &= \sigma_1\,\tau_1\,\mu_3\, 
\Phi^\dagga_{ \Cg{2y}(\bR)\,\sigma}\,.\label{C_2y}
\eal

\section{Lattice relaxation and simmetry analysis of the band structure}
\label{Lattice relaxation and band structure}
\begin{figure}
\centerline{\includegraphics[width=0.53\textwidth]{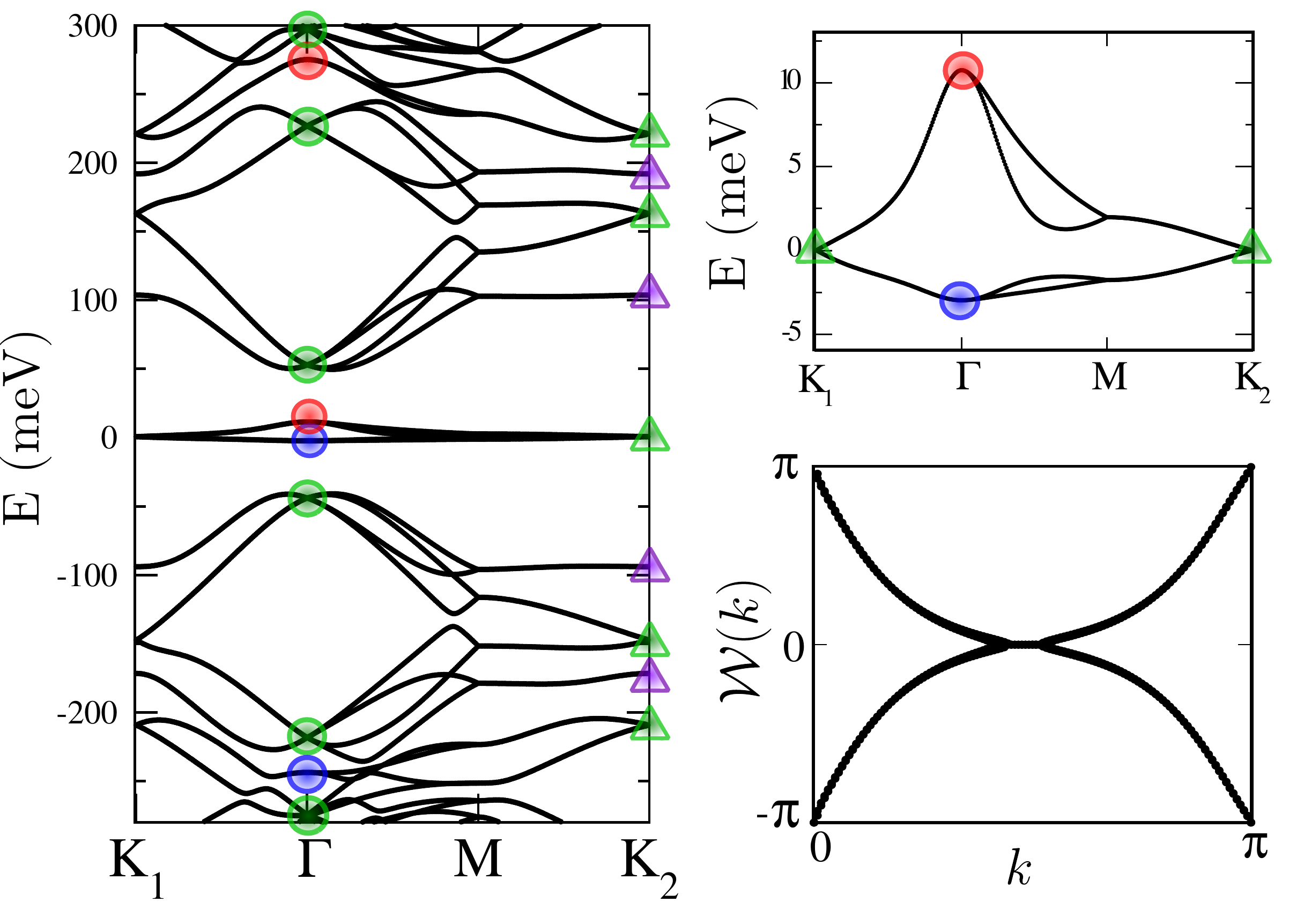}}
\caption{(a) Electronic 
band structure of twisted bilayer graphene at the angle $\theta=1.08$ after full atomic relaxation. The charge neutrality point is the zero of energy. The irreps at the $\bGamma$ point are encoded by colored circles, where blue, green and red stand for the $A_1+B_1$, $A_2+B_2$ and $E_1+E_2$ irreps of the $D_6$ space group, respectively. At the $\bK_2$ point the $E$ and $A_1+A_2$ irreps of the little group $D_3$ are represented by green and violet triangles, respectively. b) Zoom in the FBs region. c) Wilson loop of the four FBs as function of $k=G_2/\pi$.}
\label{bands}
\end{figure}

Since 
the
tBLG 
must
undergo , 
relative to the ideal superposition of two rigid graphene layers, 
a substantial lattice relaxation , whose effects have 
also been observed in recent experiments \cite{Yoo_180403806,Zhang2018,Pasupathy,Perge}, we performed lattice relaxations via classical molecular dynamics techniques 
using state-of-the-art force-fields, allowing for both in-plane and out -of-plane deformations. The details about the relaxation procedure are in Appendix \ref{AppendixA} and are essentially those in Ref.\cite{Fabrizio}.
It is well known \cite{Uchida_PRB,Dai_Nanolett,Nam_Koshino_PRB,Juricic,Yazyev,
Fabrizio,Procolo,Choi,Guinea} that, after full relaxation, the energetically less favourable AA regions shrink while the Bernal-stacked ones, AB and BA, expand in the $(x,y)$ plane. In addition, the interlayer distance along $z$
of the AA regions 
increases with respect to that of the AB/BA zones, leading to significant out-of-plane buckling deformations, genuine 'corrugations' of the graphene layers, which form protruding AA bubbles. 
The main effect of the layer corrugations is to enhance the bandgaps between the FBs and those above and below \cite{Fabrizio,Procolo,Dai_topo}, important even if
partially hindered by the h-BN encapsulation of the samples during the experiments \cite{Herrero-1}.
Moreover, since both the AB and BA 
triangular domains have expanded, the initially broad crossover region between them sharpens into narrow domain walls (DWs) that 
merge at the AA centers in the moir\'e superlattice (see \ref{lattice}). 
The electronic structure shown in Fig.~\ref{bands} is obtained with standard tight-binding calculations, see \ref{AppendixA} for further details, with the relaxed atomic positions, using hopping amplitudes tuned so to reproduce $ab$-$initio$ calculations
\cite{Trambly}. The colored circles and triangles at the 
$\bGamma$ and $\bK$ points, respectively, indicate the irreps that transform like the corresponding Bloch states. For instance, at $\bGamma$ the FBs 
consist of two doublets, the lower corresponding 
to the irreps $A_1+B_1$, and the upper to $A_2+B_2$. Right above and below the FBs, we find at $\bGamma$ two quartets, each transforming like $E_1+E_2$. At $\bK$, the FBs are degenerate and form a quartet $E+E$. Consistently with the $D_3$ little group containing $\Cg{3z}$ and $\Cg{2y}$, 
at $\bK$ we find either quartets, like at the FBs, made of degenerate pairs of doublets, each transforming like $E$, or doublets transforming like $A_1+A_2$, where $A_1$ and $A_2$ differ in the parity under $\Cg{2y}$.  
This overall doubling of degeneracies beyond their expected $D_6$ space group irreps, reflects the valley $U_v(1)$ symmetry of FBs that will be discussed below, and whose eventual breaking will be addressed later in this paper. 
We end by remarking that the so-called 'fragile' topology \cite{Bernevig_topo,Senthil-2,Vishvanath_WO,Po_PRL}, diagnosed by the odd winding of the Wilson loop (WL) \cite{Bernevig_topo}, is actually robust against lattice corrugations \cite{Dai_topo} and relaxation,  
as shown by panel (c) in Fig.~\ref{bands}.

\subsection{SU(2) symmetry and accidental degeneracy along $C_{2y}$ invariant lines}
\label{accidental}
We note that along all directions that are invariant under $\Cg{2y}$, which include the diagonals 
as well as all the edges of the MBZ, the electronic bands show a twofold degeneracy between Bloch states 
that transform differently under $\Cg{2y}$. These lines corresponds to the 
domain walls (DWs) in real space. This "accidental" degeneracy is a consequence of the interplay between $U_v(1)$ and $\Cg{2y}$ symmetries. Indeed, along $\Gamma \to K_{1,2}$ and $M \to K_{1,2}$ in the MBZ we have that:
\beal
\mC{2y}\Big(\Psi^\dagga_{\bk\sigma}\Big) = \sigma_1\,\tau_1\,\mu_3\,
\Psi^\dagga_{\bk\sigma}\,,
\eal
and similarly for $\Phi^\dagga_{\bk\sigma}$. It follows that the generator 
of $U_v(1)$, i.e., the operator $\sigma_0\,\tau_3\,\mu_0$, 
anticommutes with the expression of $\Cg{2y}$ along the lines invariant under that same symmetry, namely the operator $\sigma_1\,\tau_1\,\mu_3$, and both commute with the Hamiltonian. Then
also their product, $\sigma_1\,\tau_2\,\mu_3$, commutes with the Hamiltonian and anticommutes with the other two. The three operators 
\beal
2T_3 &= \sigma_0\,\tau_3\,\mu_0\,,&
2T_1 &= \sigma_1\,\tau_1\,\mu_3\,,&
2T_2 &= \sigma_1\,\tau_2\,\mu_3\,,\label{SU(2)}
\eal
thus realise an $SU(2)$ algebra and all commute with the Hamiltonian $\hat H_\bk$ in momentum space for any $\Cg{2y}$-invariant $\bk$-point. This 
emergent $SU(2)$ symmetry is therefore 
responsible of the degeneracy of eigenstates with opposite parity under $\Cg{2y}$.  We note that $C_{2x}$ instead commutes with $U_v(1)$, so that there is no $SU(2)$ symmetry protection against valley splitting along 
$C_{2x}$ invariant lines ($\Gamma \to M$). 
   
\section{Phonons in twisted bilayer graphene}
\label{Phonons in twisted bilayer graphene}
The $U_v(1)$ valley symmetry is an emergent one since,
despite the fact that
its generator \eqn{tau_3} does not commute with the Hamiltonian, the spectrum around charge neutrality is nonetheless  $U_v(1)$--invariant.
It is therefore not obvious to envisage a mechanism that could efficiently break it. \\
However, since the lattice degrees of freedom play an important role at equilibrium, as discussed in Section \ref{Lattice relaxation and band structure}, it is possible that they could 
offer the means
to destroy the $U_v(1)$ valley symmetry. In this Section we shall show that they indeed provide such a symmetry-breaking tool.

\subsection{Valley splitting lattice modulation: the key role of the Domain Walls}
\label{deformation}

In Section \ref{$U_v(1)$ valley symmetry} we mentioned that the valley symmetry arises because, even  
though inequivalent Dirac nodes of the two layers should be coupled to each other by the Hamiltonian 
after being folded onto the same point of the MBZ (see bottom panel in Fig.~\ref{lattice}), at small angle these
matrix elements are vanishingly small and thus the valleys are effectively decoupled.  
\begin{figure}
\centerline{\includegraphics[width=0.4\textwidth]{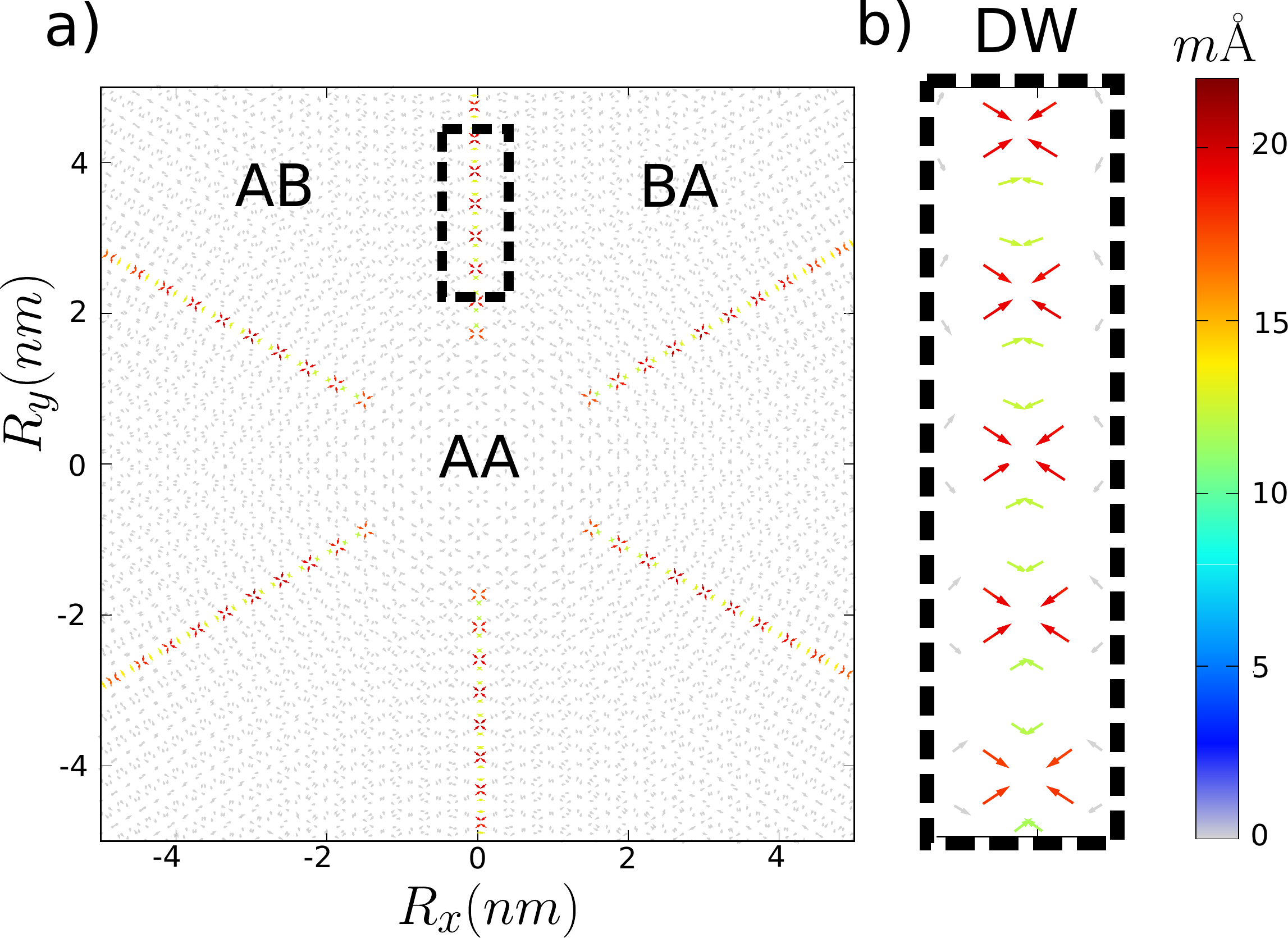}}
\centerline{\includegraphics[width=0.4\textwidth]{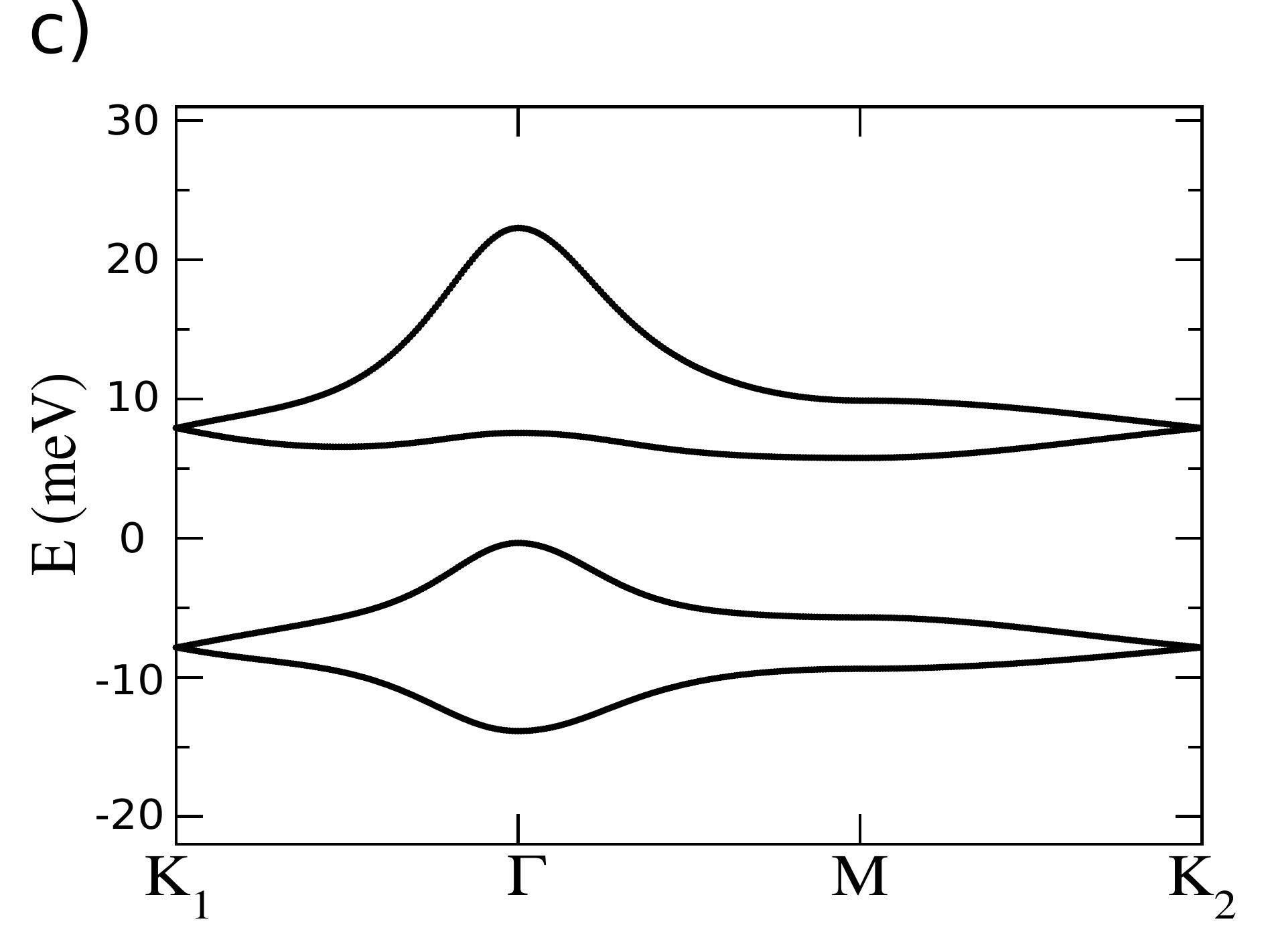}}
\caption{Panel a): the displacement  field $\boldsymbol{\eta}(\br)$, Eq.~\eqn{eta(r)}, restricted on the DWs: the direction of the atomic displacement is depicted by a small arrow, while its magnitude in  m\AA \;is expressed in colors. Panel b): displacement along the DW area highlighted by a black dashed line in panel a). The overall effect of the distortion is a narrowing of the DW. Panel c):  low energy band structure of tBLG at $\theta=1.08$ after the distortion in panel a). The two-fold degeneracies protected by the valley symmetry are completely lifted.  }
\label{def_bands}
\end{figure}
That is true if the carbon lattice, although mechanically relaxed, is unperturbed by the presence of the electrons. Once coupling with electrons is considered, we cannot exclude that for example a lattice distortion modulated with the wave vectors connecting 
the inequivalent Dirac nodes of the two layers, $K_+$ with $K'_-$ and $K'_+$ with $K_-$ in Fig.~\ref{lattice}, might instead yield a significant matrix element among the valleys. To investigate that possibility we build an ad-hoc 
distortion into the bilayer carbon atom positions. We define the vector ${\bq_{ij}}={\bK_{+,i}}-{\bK_{-,j}'}$, where $i,j=1,2,3$ run over the three equivalent Dirac points of the BZ of each layer, and the $D_6$ conserving displacement field  
\begin{equation}
\boldsymbol{\eta}(\br)= \sum_a\,\sum_{i,j=1}^3 \sin\big(\bq_{ij}\cdot\br_a\big) \,\mathbf{u}_{a,ij}\,
\delta\big(\br-\br_a\big)\,, \label{eta(r)}
\end{equation}
where $a$ runs over all atomic positions, and 
$\mathbf{u}_{a,ij}$ corresponds to a displacement of atom $a$ in direction $\bq_{ij}$, whose 
(tuneable)  magnitude is the same for all atoms. In proximity of the AA regions the distortion is locally similar to the graphene breathing mode, the in-plane transverse optical phonon at $\bK$ \cite{phonon-2} with $A_1$ symmetry.
Since by construction $\bq_{ij}$ is a multiple integer of the reciprocal lattice vectors, $\boldsymbol{\eta}(\br)$ has the same periodicity of the unit cell, i.e., the distortion is actually at the $\bGamma$ point. Moreover, the  distortion's $D_6$ invariance implies no change of space group symmetries.
\\ 

Since the most direct evidence of the $U_v(1)$ symmetry is the accidental degeneracy in the band structure
along all $\Cg{2y}$--invariant lines, corresponding just to the DWs directions in real space, we further assume 
the action of the displacement field $\boldsymbol{\eta}(\br)$ to be 
restricted to a small region in proximity of the DWs, affecting only $\approx 1\%$ of the atoms in the moir\'e supercell, see top panels in Fig.~\ref{def_bands}. The modified FBs in presence of the displacement field $\boldsymbol{\eta}(\br)$ with $\big| \mathbf{u}_{a,ij}\big| = 20$ {m\AA} are shown in the bottom panel of 
Fig.~\ref{def_bands}. Remarkably, despite the minute distortion magnitude and 
the distortion involving only the minority of carbon  atoms in the DWs, the degeneracy along $\bGamma\to \bK_1$ and $\bM\to\bK_2$ is lifted to such an extent that the four bands split into two similar copies. 
This is remarkable in two aspects. First, we repeat, because there is no space symmetry breaking. The only symmetry affected by the distortion is 
$U_v(1)$, since, by construction, $\boldsymbol{\eta}(\br)$ preserve the full space group symmetries. 
Second, the large splitting magnitude reflects an enormous strength of the effective electron-phonon coupling, whose origin is interesting. Generally speaking, in fact, broad bands involve large hoppings and large absolute electron-phonon couplings, while the opposite is expected for narrow bands. The large electron-phonon couplings which we find for the low energy bands of tBLG suggests a possible broad-band origin of the FBs, as we shall discuss in Sec.~\ref{Phonon mediated superconductivity}.  A consequence of this e-p coupling magnitude is that all potential phenomena involving lattice distortions, either static or dynamic, should be considered with a much larger priority than done so far.

\subsection{The phonon spectrum}
\label{phonon spectrum}

\begin{figure}
\centerline{\includegraphics[width=0.43\textwidth]{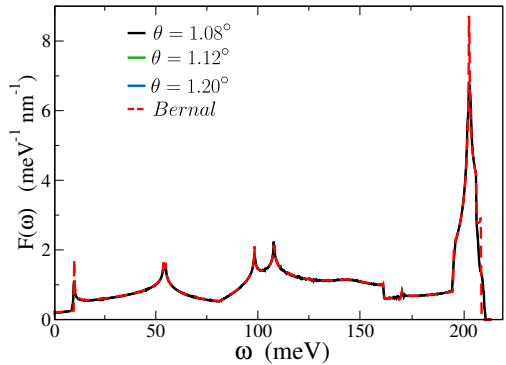}}
\caption{Phonon density of states $F(\omega)$ for Bernal stacked bilayer graphene at zero twist angle (red dashed line) and fully relaxed tBLG at $\theta=1.20,1.12,1.08$ (blue, green and black lines). }
\label{ph_dos}
\end{figure}

\begin{figure}
\centerline{\includegraphics[width=0.4\textwidth]{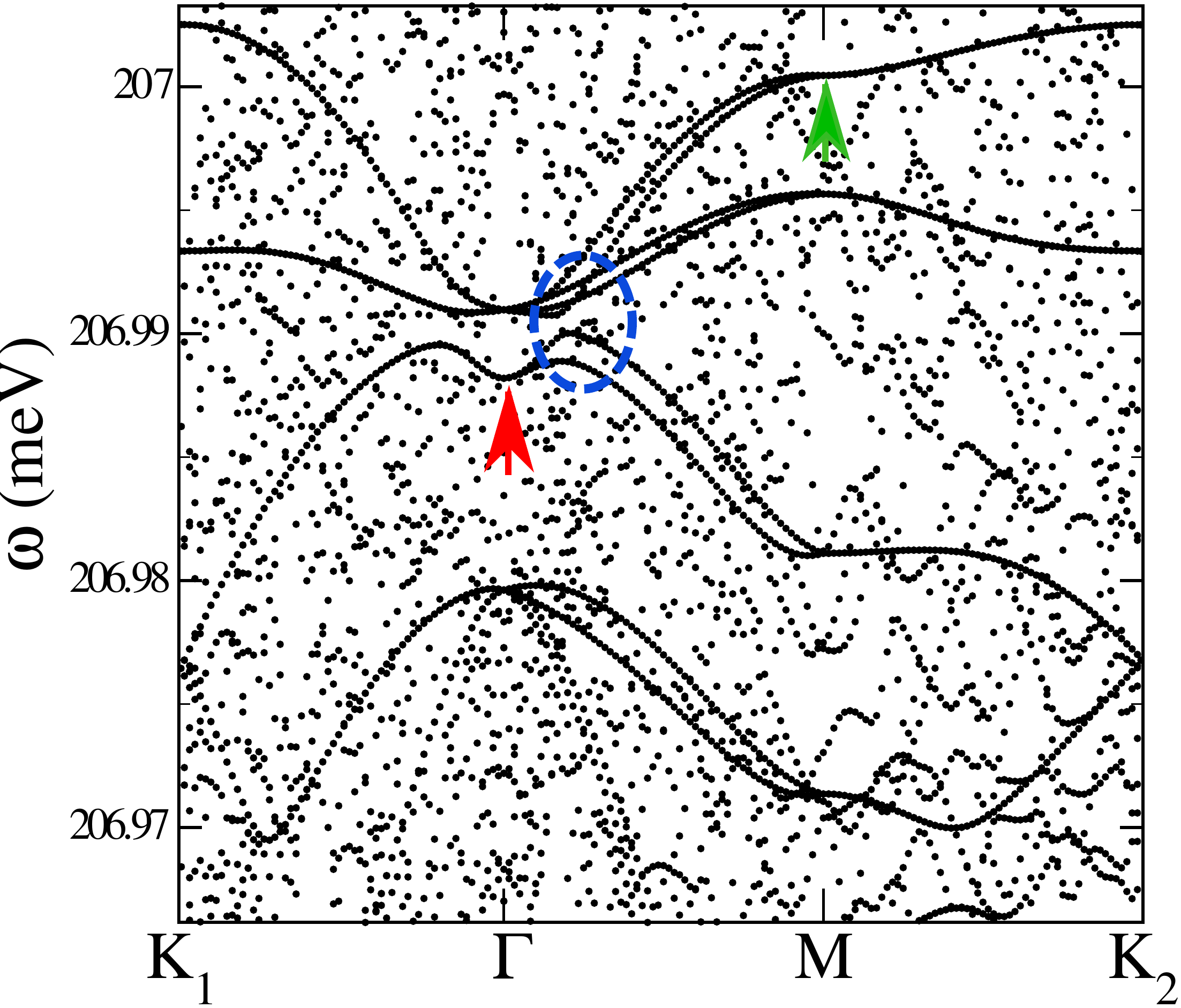}}
\caption{Zoom in the optical region of the phonon spectrum of the fully relaxed tBLG at $\theta=1.08$. Among many scattered energy levels of highly dispersive branches (not resolved in this narrow energy window) a set of 10  narrow continuous branches stands out (no line drawn through data points, which just fall next to one another). The degeneracy  of these modes is twice that expected by $D_6$  space symmetry, similar in this to electronic bands. The two-fold degenerate mode with the highest overlap with the deformation $\boldsymbol{\eta}(\br)$, drawn in  Fig.~\ref{mode_32524}, is marked by a red arrow, while the mode at $\bM$ used in section \ref{insulators} by a green arrow. The avoided crossing which occurs close to $\bGamma$ is encircled by a blue dashed line.}
\label{phonon_dispersion}
\end{figure}

We compute the phonon eigenmodes in the relaxed bilayer structure, by standard methods, see Appendix \ref{AppendixA} for details of the calculation. 
Fig.~\ref{ph_dos} shows the phonon density of states $F(\omega)$ for tBLG at three different twist angles in comparison with the Bernal AB-stacked bilayer.
As previously reported \cite{Choi,Cocemasov}, $F(\omega)$ is almost independent of the twist angle, 
which only affects the inter-layer van der Waals forces, much weaker than the in-plane ones arising from the 
stiff $C-C$ bonds. 
As a consequence, phonons in tBLG are basically those of the Bernal stacked bilayer.
This is true except for a small set of special phonon modes, clearly distinguishable 
in Fig.~\ref{phonon_dispersion} that depicts the phonon spectrum zoomed in a very narrow energy region $\approx 0.04$ $meV$ around 
the high frequency graphene $\bK$-point 
peak of the phonon density of states. Specifically, within the large number of energy levels of all other highly dispersive phonon bands, unresolved in the narrow 
energy window, a set of 10 almost dispersionless modes emerges. We note that these special modes show the same accidental degeneracy doubling along the $C_{2y}$ invariant lines as that of 
the electronic bands around the charge neutrality point. 

\begin{figure}
\centerline{\includegraphics[width=0.4\textwidth]{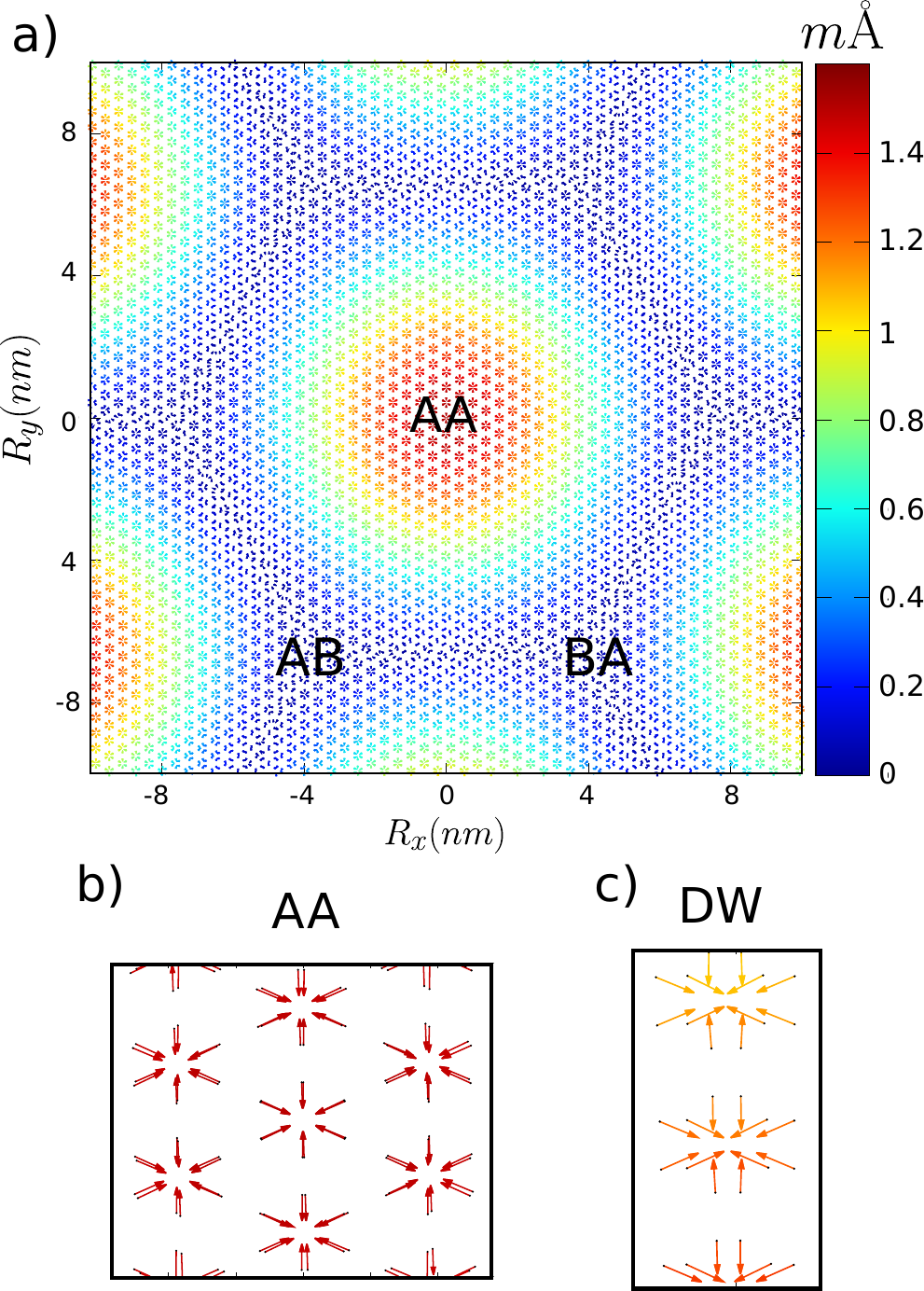}}
\caption{Panel a): atomic displacements on one of the two layers corresponding to the $A_1$-symmetry moir\'e mode of the phonon doublet marked by a red arrow in Fig.~\ref{phonon_dispersion}. The direction of displacement is represented by a small arrow centered at each atomic position, while its modulus is encoded in colors. The mean displacement per atom is $0.57$ m\AA. Panel b):  zoom in the center of an AA region, shown for both layers. Panel c): zoom along one of the 
domain walls. Note the similarity with the ad-hoc displacement in 
Fig.~\ref{def_bands}.}
\label{mode_32524}
\end{figure}

\begin{figure}
\centerline{\includegraphics[width=0.53\textwidth]{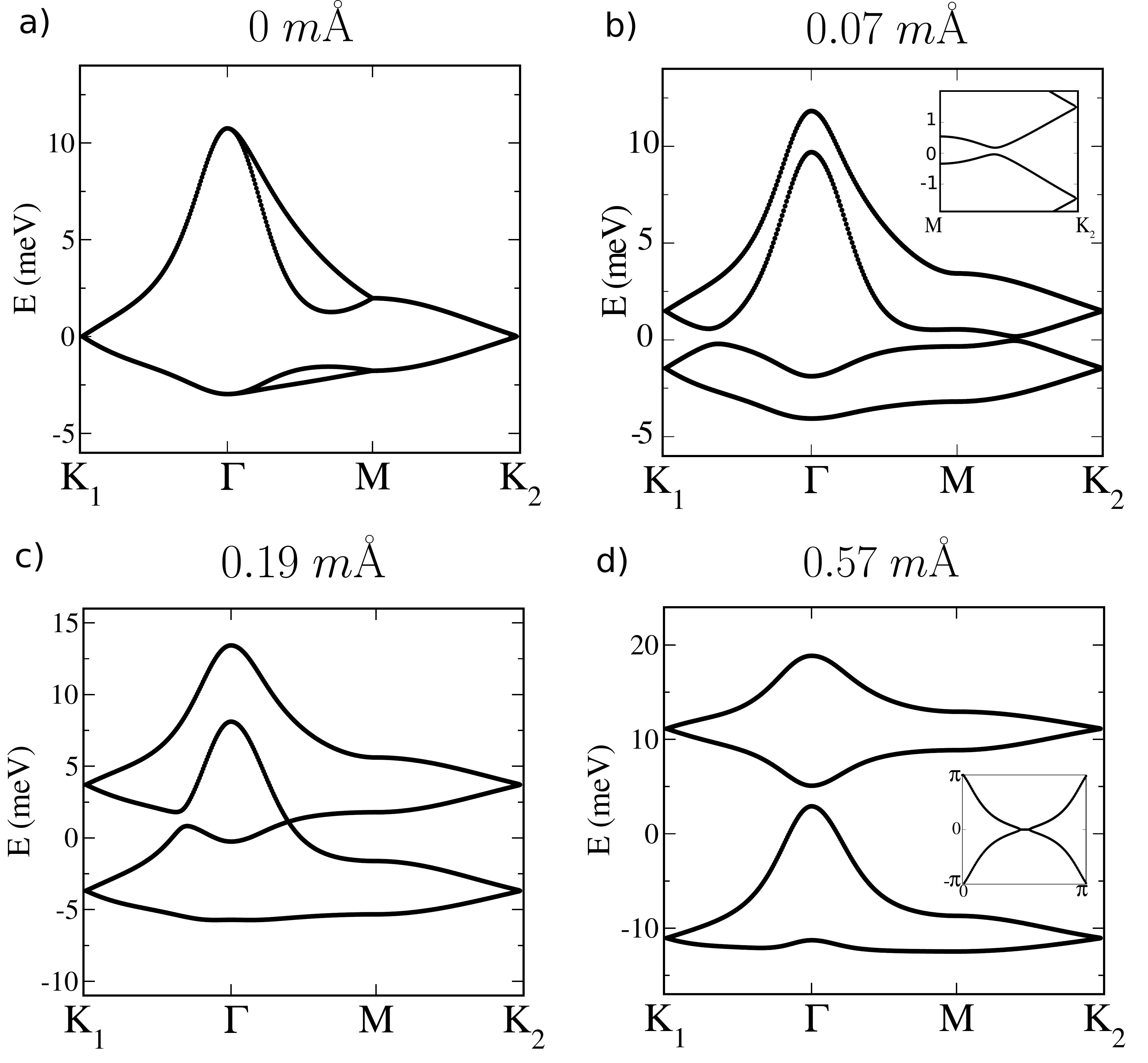}}
\caption{Evolution of the FBs when the lattice is distorted with increasing intensity along one of the two modes indicated by a red arrow in Fig.~\ref{phonon_dispersion}. 
Panel (a) undistorted. Panel (b) mean displacement per atom 0.07~m\AA. In the inset we show the avoided crossing along $\bM\to\bK_2$. Panel (c)  
mean displacement 0.19~m\AA. Now the avoided crossing appears as a genuine 
crossing protected by $\Cg{2x}$ symmetry along $\bGamma\to\bM$, which actually leads to Dirac 
points. Panel (d) mean displacement 0.57 m\AA,  
opening a gap between the FBs. Inset: Wilson loop of the lowest two bands in panel d).}
\label{frozen_phonon_bands}
\end{figure}

Similar to the electronic degeneracy, whose underlying  $U_v(1)$ symmetry
arises from vanishingly weak hopping matrix elements between interlayer K-K' points in the Hamiltonian, the mechanically weak van der Waals interlayer coupling here leads to an effective $U_v(1)$ symmetry for this group of lattice vibrations.  Their poor dispersion 
is connected with  a displacement which is non-uniform in the supercell, and is strongly modulated on the moir\'e length scale, a distinctive feature of these special modes that we shall denote as `moir\'e phonons'.  
In particular, it is maximum in the center of the AA zones, finite in the DWs,  and negligible in the large AB and BA Bernal regions. 
The overlap between the displacement $\boldsymbol{\eta}(\br)$ in \eqn{eta(r)} and the $33492$ phonons of the $\theta=1.08$ tBLG at the $\bGamma$ point is non-negligible 
only for those moir\'e phonons. In particular, we find the highest overlap with the 
doubly degenerate mode marked by a red arrow in Fig.~\ref{phonon_dispersion}, and which 
transforms like $A_1+B_1$. In Fig.~\ref{mode_32524}a) we show the real 
space distortion corresponding to the $A_1$ component of the doublet, where the displacement direction 
is represented by small arrows, while its intensity is encoded in colors.
This  inspection of the eigenvectors of these modes at the atomistic level reveals a definite 
underlying
single-layer graphene character, specifically that of the $A_1$ symmetry transverse optical mode at $\bK$.  
Since the graphene $\bK$-point does \textit{not} fold into the bilayer $\Gamma$-point, their appearance 
along the whole $\Gamma\to\bK$ line in the spectrum of the fully relaxed bilayer must be merely a consequence of relaxation, a relaxation that is 
particularly strong precisely in the AA and DW regions. 

\section{Insulating state at charge neutrality}
 
\begin{figure}
\centerline{\includegraphics[width=0.37\textwidth]{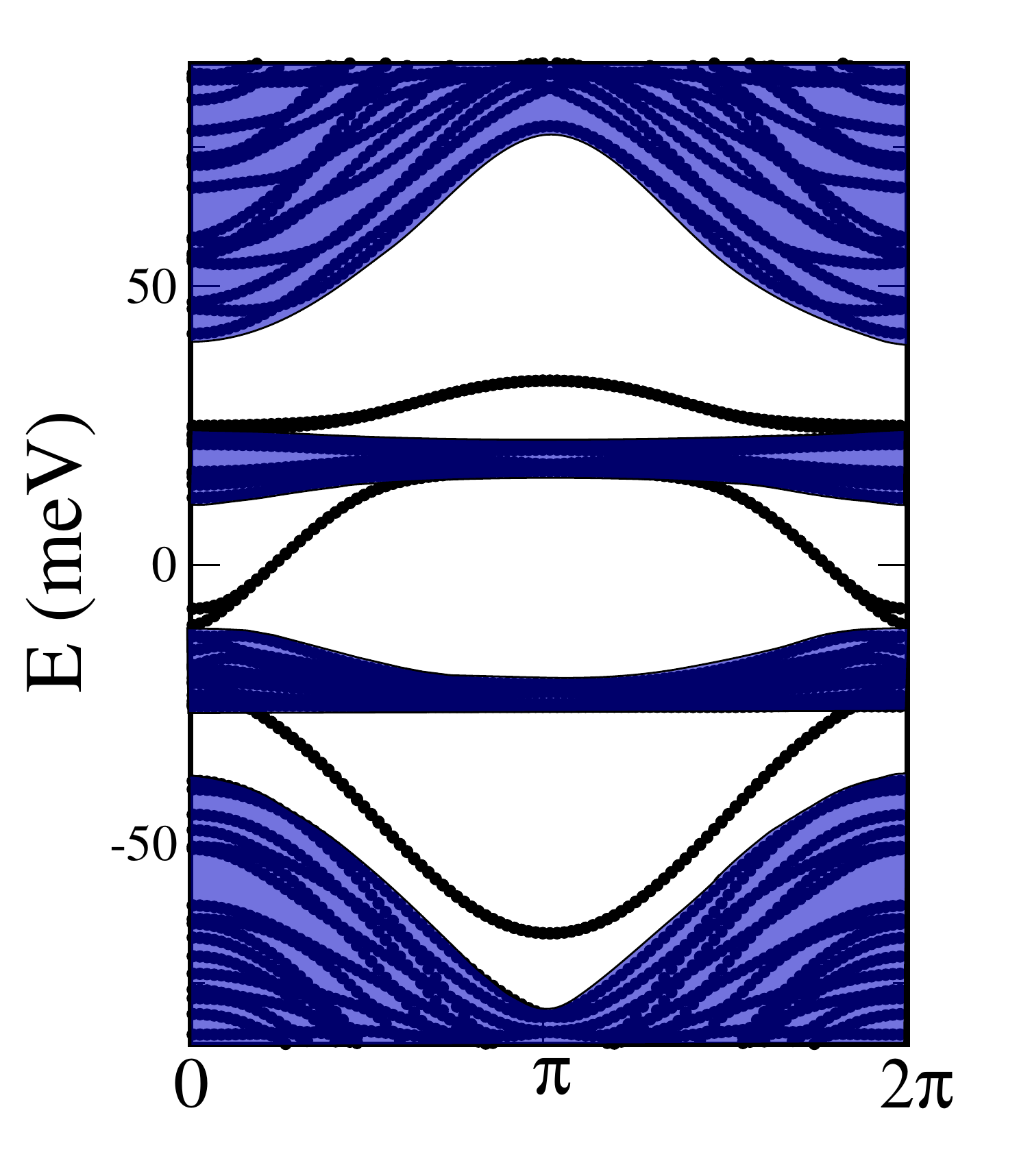}}
\caption{The band structure of twisted bilayer graphene at $\theta\approx 1.08^\circ$
in a ribbon geometry with open boundary conditions. The atomic structure inside the unit cell  (replicated 7 times along the $x$ direction) has been deformed with the $A_1$ symmetric moir\'e phonon mode. With a mean deformation amplitude of $\approx 1.14\; m$\AA, that mode is so strongly coupled to open a large electronic gap at CNP of $\approx 25$ $meV$. The bulk bands have been highlited in blue to emphasize the presence of edge states both within the phonon-driven FBs gap and the gaps above and below them.
}
\label{edge states}
\end{figure}

We will next focus on the effect on the electronic band structure of 
a carbon atom displacement corresponding to the two degenerate phonon modes $A_1$ and  $B_1$ at $\Gamma$, which should affect the valley symmetry as the displacement in Fig.~\ref{def_bands}.   
In order to verify that,  we carried out a frozen phonon calculation of the modified FB electronic 
structure with increasing intensity of the deformation \cite{supp_video}.  
Remarkably, despite transforming as different irreps ($A_1$ or $B_1$), both frozen phonon distortions
are not only degenerate, but have exactly the same effect on the bands.
 As soon as the lattice is distorted, see Fig.~\ref{frozen_phonon_bands} b), the fourfold degeneracy at 
 $\bK_1$ and $\bK_2$, and the twofold one at $\Gamma$ and $\bM$ 
is lifted, and small avoided crossings appear and start to move from $\bK_{1(2)}$ towards the $\bM$ points. Once they cross $\bM$, they keep moving along $\bM\to\bGamma$, 
see Fig.~\ref{frozen_phonon_bands} c). 
However, along these directions, the $C_{2x}$ symmetry prevents the avoided crossings, and thus 
leads to six elliptical Dirac cones. Finally,  once they reach $\bGamma$ at a threshold value of the distortion,  
the six Dirac points annihilate so that a gap opens at charge neutrality, 
see Fig.~\ref{frozen_phonon_bands} d).
This gap-opening mechanism is very efficient, with large splittings even for small values of the atomic displacement amplitude shown in Fig.~\ref{frozen_phonon_bands}. 
For instance, an average displacement as small as $\approx 0.5~\text{m\AA}$ per atom is enough to completely separate the four FBs and to open a gap at charge neutrality.\\
We emphasize that this occurs without breaking any spatial symmetry of the tBLG, just $U_v(1)$. 
As a consequence, the insulator state 
possesses a non trivial topology, as highlighted by the odd-winding of the Wilson loop of the lowest two bands, shown in the inset of the same figure 
\ref{frozen_phonon_bands} d). 
In turn, the non trivial topology of the system implies
the existence of edge states within the gap separating the two lower flat bands from the two upper ones. We thus 
recalculated the band structure  
freezing the moir\'e phonon with $A_1$ symmetry at $\bGamma$ in a ribbon geometry, which is obtained by cutting the tBLG along two parallel domain walls in the 
$y$-direction at a distance of 7 supercells. The ribbon has therefore translational symmetry along $y$, but it is confined in the $x$-direction. 
In Fig.~\ref{edge states} we show the single-particle energy levels as function of the momentum $k_y$ along the $y$-direction. 
Edge states within the gap at charge neutrality are clearly visible. In particular, we find for each edge two counter-propagating modes. 

As a matter of fact, recent experiments do report the existence of a finite gap also at charge neutrality \cite{Efetov,Eva-Andrei}, which is actually bigger than at other non zero integer fillings, and appears without a manifest breakdown of time reversal symmetry T \cite{Efetov} or $\Cg{2z}$ symmetry \cite{Eva-Andrei}, and with the FBs still well separated from other bands \cite{Eva-Andrei}. These evidences seem not to support an interaction driven gap, which 
would entail either T or $\text{T}\,\Cg{2z}$ symmetry breakings 
\cite{Efetov, Vishwanath-charge-neutrality}, or else the FBs touching other bands at the $\bGamma$ point 
\cite{Efetov}.  Our phonon-driven insulator at charge neutrality breaks instead $U_v(1)$ and, eventually, $\Cg{2y}$ 
if the frozen phonon has $B_1$ character, both symmetries experimentally elusive. However, the edge states that we predict could be detectable by STM or STS, thus providing support or disproving the mechanism that we uncovered. \\
We end mentioning that near charge neutrality there are compelling evidences of a substantial 
breakdown of $\Cg{3z}$ symmetry \cite{Pasupathy,Perge,Eva-Andrei}, which, although is not expected to stabilise on its own an insulating gap unless the latter were in fact just a pseudo gap \cite{Vishwanath-charge-neutrality}, still it is worth being properly discussed, which we postpone to Section \ref{C3z symmetry breaking}.
\subsection{$\text{E}\otimes\text{e}$ Jahn-Teller effect}
\label{Jahn Teller}
The evidence that the two degenerate modes marked by a red arrow in Fig.~\ref{phonon_dispersion} 
produce the same band structure in a frozen-phonon distortion
is reminiscent of a $\text{E}\otimes\text{e}$ Jahn-Teller effect, i.e., the coupling of a doubly degenerate vibration with a doubly degenerate electronic state \cite{Englman}. \\
Let us consider the action of the two $\Gamma$-point  $A_1$ and $B_1$ phonon modes, hereafter denoted as $q_1$ and $q_2$, along the $\Cg{2y}$ invariant lines. 
We note that the $A_1$ mode,  $q_1$, although invariant under the $D_6$ group elements,  is able to split the degeneracy along those lines. Therefore it must be coupled to the electrons through a $D_6$ invariant operator that does not commute with the $U_v(1)$ generator $\tau_3$. That in turn cannot but coincide with 
$\Cg{2y}$ itself, which, along the invariant lines, is the operator $T_1=\sigma_1\,\tau_1\,\mu_3/2$ in Eq.~\eqn{SU(2)}. On the other hand, 
the $B_1$ mode, $q_2$, is odd under $\Cg{2y}$, and thus it must be associated with an operator that anticommutes with $\Cg{2y}$ and does not commute with $\tau_3$. 
The only possibility that still admits a $U(1)$ valley symmetry is the operator $T_2=\sigma_1\,\tau_2\,\mu_3/2$ in Eq.~\eqn{SU(2)}. Indeed, 
with such a choice, the electron-phonon Hamiltonian is
\beal
H_\text{el-ph} &= -g\,\Big( q_1\,T_1 + q_2\,T_2\Big)\,,\label{Exe-Ham}
\eal
with $g$ the coupling constant. This commutes with the operator 
\beal
J_3 &= T_3 + L_3 = \fract{\tau_3}{2} +\bq\wedge\mathbf{p}\,,\label{J3}
\eal
(where $\mathbf{p} = (p_1,p_2)$ the conjugate variable of the displacement $\bq=(q_1,q_2)$), 
the generator of a generalized $U_v(1)$ symmetry that involves electron 
and phonon variables. As anticipated, the 
Hamiltonian \eqn{Exe-Ham} describes precisely 
a $\text{E}\otimes\text{e}$ Jahn-Teller problem~\cite{Englman, Airoldi-PRB1996, JTMott-PRB1997}. \\
Since the phonon mode $\bq$ is almost dispersionless, see Fig.~\ref{phonon_dispersion}, we can think of it 
as the vibration of a moir\'e supercell, as if the latter were a single, though very large, molecule, and the tBLG a molecular conductor. In this language, the band structures shown in Fig.~\ref{frozen_phonon_bands} 
would correspond to a static Jahn-Teller distortion. 
However, since the phonon frequency is substantially larger 
than the width of the flat bands, we cannot exclude the possibility of a dynamical Jahn-Teller effect that 
could mediate superconductivity \cite{Airoldi-PRB1996, ExeSC-PRL2004}, or 
even stabilize a Jahn-Teller Mott insulator \cite{JTMott-PRB1997} in presence of a strong enough interaction. \\
The Jahn-Teller nature of the electron-phonon coupling entails 
a very efficient mechanism to split the accidental degeneracy, linear in the displacement within a frozen phonon calculation. However, this does not explain why an in-plane displacement as small as {0.57~m\AA} is able to split the formerly degenerate states at $\bGamma$ by an amount as large as $ 15~\text{meV}$, see
Fig.~\ref{frozen_phonon_bands} d), of the same order as the original width of the flat bands.  To clarify that, we note that this displacement would yield a change in 
the graphene nearest neighbour hopping of around $\delta t \simeq 3.6~\text{meV}$, see Eq.~\eqn{A1}, which in turn entails a splitting at $\bGamma$ of $6\,\delta t
\sim 21~\text{meV}$, close to what we observe. We believe that such correspondence is not accidental, but  indicates that the actual energy scale underneath the flat bands is on the order of the bare graphene bandwidth, rather than the flat bandwidth itself. We shall return on this issue later in Section \ref{Phonon mediated superconductivity}.

\section{Insulating states at other 
commensurate fillings}
\label{insulators}
\begin{figure}
\centerline{\includegraphics[width=0.4\textwidth]{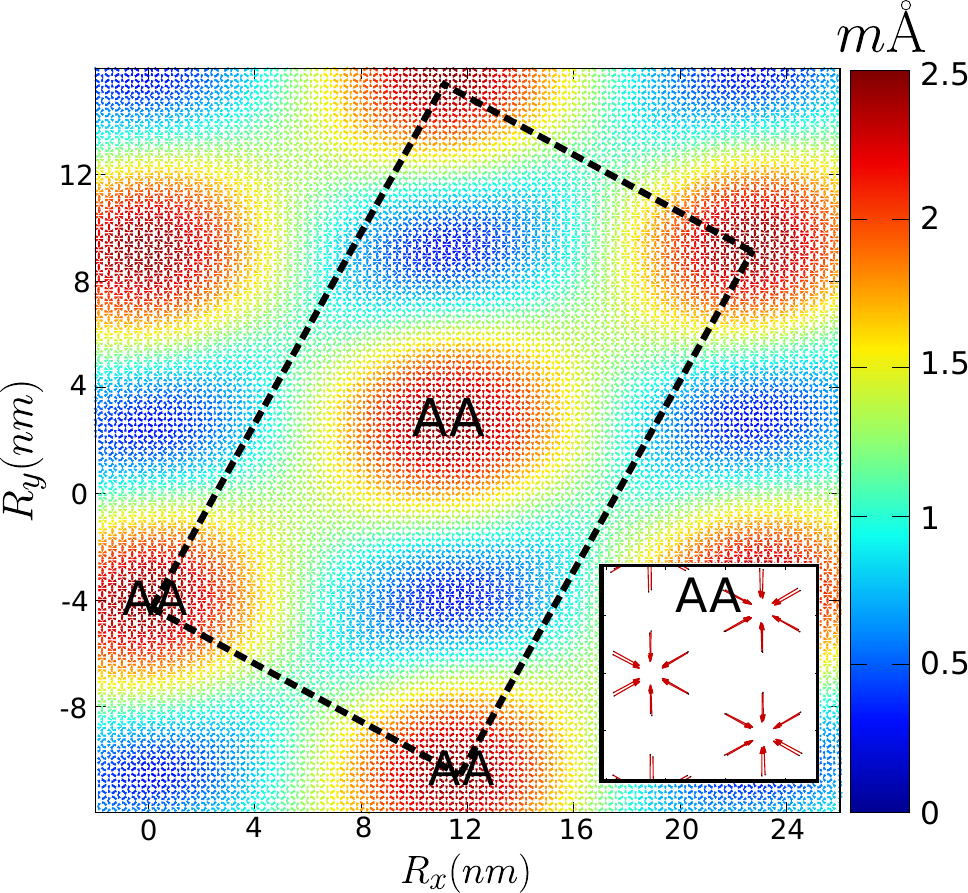}}
\caption{Atomic displacements on one of the two layers corresponding to one of the moir\'e modes at $M$  marked by a green arrow in Fig.~\ref{phonon_dispersion}. The direction of displacement is represented by a small arrow centered at each atomic position, while its modulus is encoded in colors. The mean deformation is $1.8~\text{m\AA}$ and leads to the DOS in Fig.~\ref{DOS_M}. The inset shows a zoom in the AA region close to the origin. The rectangular unit cell, now containing twice the number of atoms,
is highlighted by a black dashed line.  }
\label{phonon_M}
\end{figure}

\begin{figure}
\centerline{\includegraphics[width=0.45\textwidth]{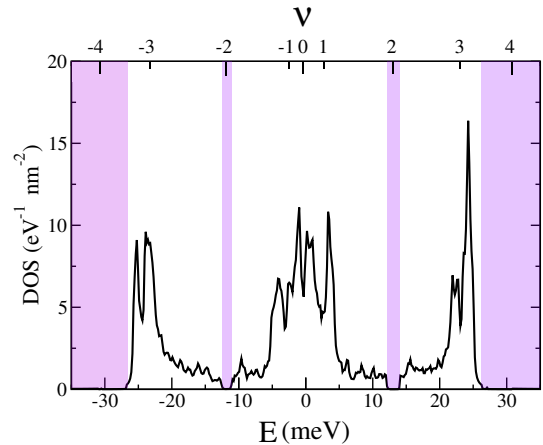}}
\caption{Density of states in the flat bands region after the tBLG has been distorted by one of the two degenerate modes at $\bM$ marked by a green arrow in Fig.~\ref{phonon_dispersion}. A mesh of 
$40\times40$ points in the MBZ has been used. The DOS is expressed as function of energy, 
the zero corresponding to the CNP, 
and corresponding occupancy $\nu$ of the FBs.
The band gaps at $\nu=\pm 2,\pm 4$ are highlighted in violet. The mean displacement per atom  
is $1.8~\text{m\AA}$.}
\label{DOS_M}
\end{figure}

\begin{figure}
\centerline{\includegraphics[width=0.45\textwidth]{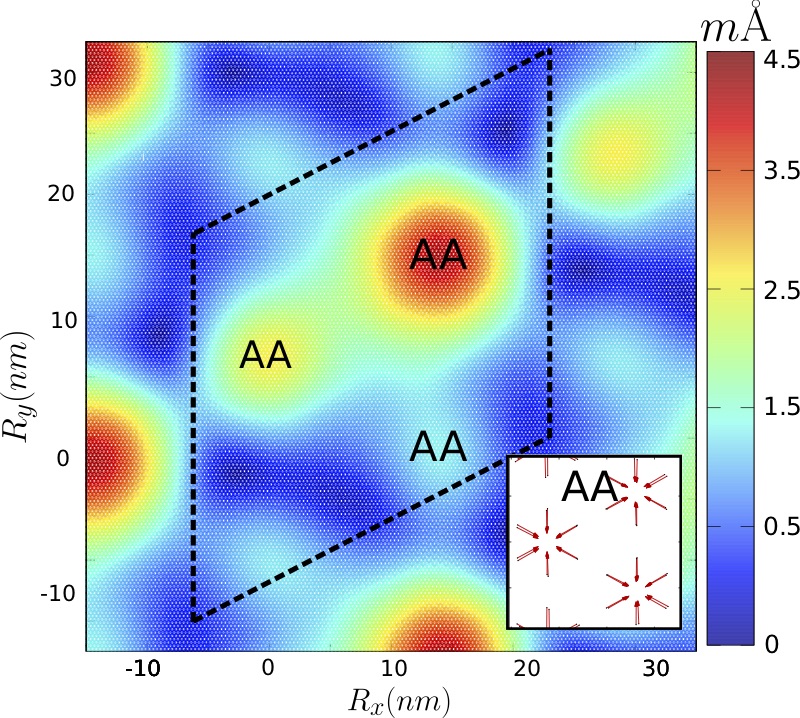}}
\caption{Atomic displacements on one of the two layers corresponding to multicomponent deformation obtained combining moir\'e JT phonons at three inequivalent $\bM$ points. The mean deformation is $1.7~\text{m\AA}$ and leads to the DOS in Fig.~\ref{DOS_M4}. The intensity of displacement is encoded in colors. The inset shows a zoom in one of the AA regions when both layers are considered. The unit cell, denoted by a black dashed line, is four times larger than the original one. }
\label{phonon_M4x}
\end{figure}

\begin{figure}
\centerline{\includegraphics[width=0.45\textwidth]{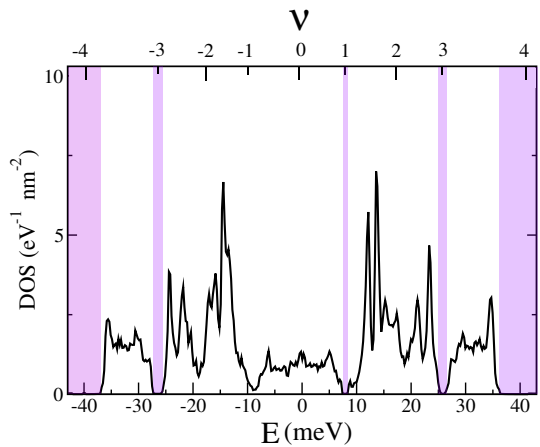}}
\caption{Electronic density of states  with a multicomponent lattice distortion obtained by freezing 
a combination of the modes at the three inequivalent $\bM$ points. Band gaps at $\nu=+1,\pm 3,\pm 4$ are highlighted in violet.  The mean displacement per atom is $1.7~\text{m\AA}$. }
\label{DOS_M4}
\end{figure}

The $\Gamma$-point distortion described above can lead to an insulating state at charge neutrality, possibly connected with the insulating state very recently reported~\cite{Efetov,Eva-Andrei}.
The same phonon branch might also stabilize insulating states at other integer 
occupancies of the mini bands besides charge neutrality. However, this necessarily requires freezing 
a mode at a high-symmetry $\bk$-point different from $\bGamma$ in order to get rid of the  
band touching at the Dirac points, $\bK_1$ and $\bK_2$, protected by the $\Cg{6z}$ symmetry. 
We consider one of the two degenerate phonon modes at the $\bM$-point marked by a green arrow in Fig.~\ref{phonon_dispersion}.
This mode has similar features as the Jahn-Teller one at $\bGamma$, even if it belongs to
an upper branch due to an avoided crossing along $\bGamma\to\bM$ (blue dashed line in Fig.~\ref{phonon_dispersion}).
In Fig.~\ref{phonon_M} we depict
this mode, which still transforms as the $A_1$ of graphene $\bK$-point 
on the microscopic graphene scale, but whose long-wavelength modulation now forms a series of ellipses elongated along some of the DWs, thus macroscopically breaking the $C_{3z}$ symmetry of the moir\'e superlattice.\\
In Fig.~\ref{DOS_M} we show the DOS of the FBs obtained by a frozen-phonon realistic tight-binding calculation. Besides the band gaps at $\nu= \pm 4$, which separates the FBs from the others bands, now small gaps opens at $\nu=\pm 2$ with an average atomic displacement of $1.8~\text{m\AA}$ induced 
by the mode at $\bM$. \\
Finally we also considered a more exotic multi-component distortion induced by a combination of the modes at the three inequivalent $\bM$ points, which quadruples the unit cell (see Fig.~\ref{phonon_M4x}). The resulting DOS of the FBs is shown in Fig.~\ref{DOS_M4}, and 
displays small gaps at the odd integer occupancies $\nu=1$ and $\nu=\pm 3$. 

The very qualitative  conclusion of this exploration is that frozen phonon distortions with various $\bk$-vectors can very effectively yield Peierls-like insulating states at integer hole/electron fillings.
Of course all distortions at $\bk$-points different from $\Gamma$ also represent super-superlattices, with an enlargement of the unit cell that should be verifiable in such a Peierls state, 
even if
the displacement is 
a tiny fraction of the equilibrium C--C distance. We 
note here 
that the 
the zone boundary phonons are less effective in opening gaps at non zero integer fillings than the zone center phonons are at the charge neutrality point. The reason is that away from charge neutrality the Jahn-Teller effect alone is no longer sufficient;  one needs to invoke zone boundary phonons that enlarge the unit cell, and thus open gaps at the boundaries of the folded Brillouin zone. The efficiency of such gap opening mechanism is evidently lower than that of the Jahn-Teller $\bGamma$-point mode. 

\subsection{$\Cg{3z}$ symmetry breaking}
\label{C3z symmetry breaking}
We observe that the multicomponent distortion with the phonons frozen at the 
inequivalent $\bM$ points leads to a width of the FBs five times larger than 
in the undistorted case, see Fig.~\ref{DOS_M4} as opposed to  Fig.~\ref{frozen_phonon_bands}~a), which is not simply a consequence of the tiny gaps that open at the boundary of the reduced Brillouin zone. Such substantial bandwidth increase suggests that the tBLG may be intrinsically unstable to $\Cg{3z}$ symmetry breaking, especially near charge neutrality. Electron-electron interaction treated in mean-field does a very similar job \cite{Perge}. Essentially, 
both interaction- and phonon-driven mechanisms act right in the same manner: they move the two van Hove singularities of the fully symmetric band structure away from each other, and split them into two, with the net effect of increasing the bandwidth. As such, those two mechanisms will cooperate to drive the C$_{3z}$ symmetry breaking, or enhance it when explicitly broken by strain, not in disagreement with experiments \cite{Pasupathy,Perge,Eva-Andrei}. The main difference is that the moir\'e phonons also break $U_v(1)$, and thus are able to open gaps at commensurate fillings that $\Cg{3z}$ symmetry breaking alone would not do.     
\section{Phonon mediated superconductivity}

\label{Phonon mediated superconductivity}

\begin{figure}
\centerline{\includegraphics[width=0.45\textwidth]{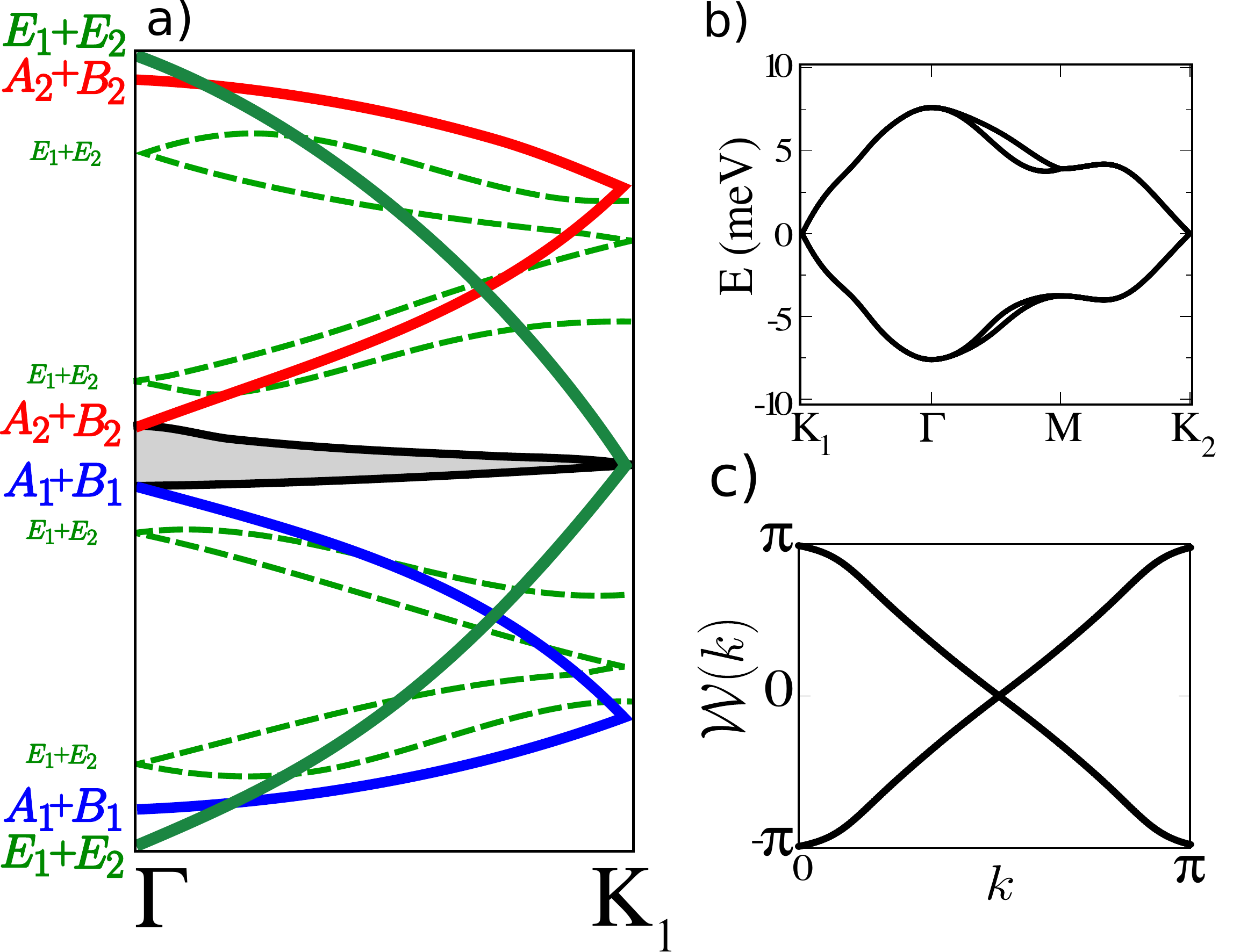}}
\caption{Panel a): sketch of the elementary band representation along $\bGamma\to\bK_1$ taking into account 
the 32 bands closest to the charge neutrality point and without allowing avoided crossing between same symmetry Bloch states. Blue, red and green lines refer to Bloch states that at $\bGamma$ transform like 
$A_1+B_1$, $A_2+B_2$ and $E_1+E_2$, respectively. Should we allow for avoided crossings, close to charge neutrality we would 
obtain the four flat bands shown as black lines surrounding the shaded region. 
Panel b): the FBs obtained by a model tight-binding Hamiltonian that includes only the solid bands 
of panel a). The details of this Hamiltonian are given in Appendix~\ref{AppendixB}. Panel c): Wilson loop corresponding to the four bands in panel b) fully occupied.}
\label{sketch}
\end{figure}

Here we indicate how the phenomena described above can connect to a superconducting state mediated by electron-phonon coupling. For that, we need to build a minimal tight-binding model containing a limited set of orbitals. For the sake of simplicity we shall not require the model Hamiltonian to reproduce precisely the shape of all the bands around charge neutrality, especially those above or below the FBs, but only the correct elementary band representation, topology and, obviously, the existence of the four flat bands separated from the all others. Considering 
for instance only the 32 states at $\bGamma$ closest to the charge neutrality point, and maintaining 
our assumption of WOs centred at the Wyckoff positions $2c$, those states (apart from avoided crossings allowed by symmetry) 
would evolve from $\bGamma$ to $\bK_1$ in accordance with the $D_6$ space group 
as shown in Fig.~\ref{sketch}. Once we allow same-symmetry Bloch states to repel each other along $\bGamma\to\bK_1$, the band representation can
look similar to the real one (Fig.~\ref{bands}), including the existence of the four FBs that 
start at $\bGamma$ as two doublets, $A_1+B_1$ and $A_2+B_2$, and end at $\bK_1$ 
as two degenerate doublets, each transforming as the 2d-irrep $E$, see the two solid black lines in 
Fig.~\ref{sketch}. While this picture
looks compatible with the actual band structure, we shall take a further simplification and just consider the thicker red, blue and green bands in Fig.~\ref{sketch}, 
which could still produce flat bands with the correct symmetries. This oversimplification obviously implies giving up the possibility to accurately reproduce the shape of the FBs - but it makes the algebra much simpler. 

Within this approximation the components of the spinor operators in Eq.~\eqn{spinors} are actually single fermionic operators, so that we limit ourselves to just four WOs for each sublattice, AB or BA, 
and valley, 1 and 2. Two of such WOs transform like the 1d-irreps,  one even, $A_1$, and the other odd, $A_2$, under $\Cg{2x}$. The other two instead transform like the 2d-irrep $E$.  We build a minimal tight-binding model 
\be
H_\text{el} = -\Delta\,\sum_{\bR\sigma}\,\Psi^\dagger_{\bR\sigma}\sigma_0\tau_0\mu_3
\Psi^\dagga_{\bR\sigma} +T_\text{el},\label{Ham}
\ee
where $\Delta$ splits the $s$ from the $p$ WOs of the 1d-irreps, and $T_\text{el}$ 
includes first and second neighbour hopping between AB and BA regions in the moir\'e superlattice 
compatible with all symmetries, see Appendix \ref{AppendixB} for details. 
In Fig.~\ref{sketch} we show the resulting FBs as well as their Wilson loop.  
We emphasise that the FBs arise in this picture from a sequence of avoided crossings between a large set of relatively broad bands that strongly repel each other away from the high symmetry points, rather than from 
truly localized' WOs. 
This mechanism is also compatible with, 
and in fact  behind, 
the strong 
electron-phonon coupling strength, and the  consequently large
effects on the FBs of the Jahn-Teller phonons that, according to Eq.~\eqn{Exe-Ham}, simply modulate the AB--BA hopping.

\subsection{Mean-field  superconducting state}
\label{ph_BCS}
\begin{figure}[t]
\vspace{-1.5cm}
\centerline{\includegraphics[width=0.5\textwidth]{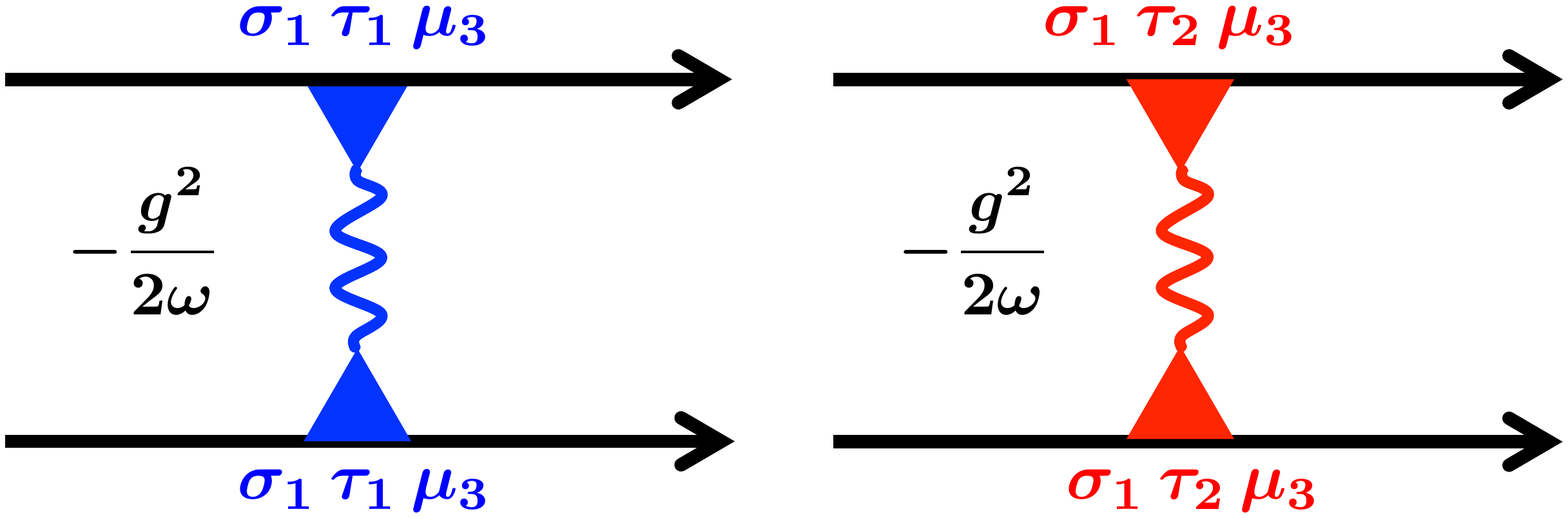}}
\vspace{-1.8cm}
\caption{Phonon mediated attraction. The two scattering channels corresponds to the two phonons and have the same amplitude $g^2/2\omega$. }
\label{scattering}
\end{figure}

Neglecting the extremely small dispersion of the moir\'e Jahn-Teller phonons, we can write their 
Hamiltonian simply as  
\be
H_\text{ph} = \fract{\omega}{2}\,\sum_\bR\, \Big(\,
\bp_\bR\cdot\bp_\bR + \bq_\bR\cdot\bq_\bR
\,\Big)\,,\label{Ham-ph}
\ee 
with $\omega\simeq 207~\text{meV}$.\\
Rather than trying to model more faithfully the Jahn-Teller coupling \eqn{Exe-Ham}, we 
shall follow a simplified approach based just on symmetry considerations. \\
In general, we could integrate out the phonons to obtain a retarded electron-electron attraction that can mediate superconductivity. 
However, since here the phonon frequency is much larger that the bandwidth of the FBs, where the chemical potential lies, we can safely neglect retardation effects making a BCS-type approximation virtually exact. 
The attraction thus becomes instantaneous and can be represented as in Fig.~\ref{scattering}.
The phonon couples electrons in 
nearest neighbor AB and BA regions, giving rise to an inter-moir\'e site spin-singlet pairing, a state which we expect to be much less affected by Coulomb repulsion than an on-site one. 
Therefore, neglecting Coulomb repulsions we can concentrate on the pairing channel between nearest neighbor AB and BA regions. The scattering processes in Fig.~\ref{scattering} imply that the pairing channels are only $\tau_1\,\mu_0$ and $\tau_1\,\mu_3$, corresponding to inter valley pairing, 
as expected because time reversal interchanges the two valleys.\\
Having assumed pairing between nearest neighbor AB and BA regions, 
we must identify pair functions in momentum space that connect nearest neighbor unit cells, and transform properly under $\Cg{3z}$.
These functions 
are  
\beal
\gamma(\bk) &= \esp{i\bk\cdot(\bba+\bbb)/3}\,
\bigg(1+ \esp{-i\bk\cdot\bba} + \esp{-i\bk\cdot\bbb}\,\bigg)\,,\\
\gamma_{+1}(\bk) &= \esp{i\bk\cdot(\bba+\bbb)/3}\,
\bigg(1+ \omega\,\esp{-i\bk\cdot\bba} + \omega^*\,\esp{-i\bk\cdot\bbb}\,\bigg)\,,\\
\gamma_{-1}(\bk) &= \esp{i\bk\cdot(\bba+\bbb)/3}\,
\bigg(1+ \omega^*\,\esp{-i\bk\cdot\bba} + \omega\,\esp{-i\bk\cdot\bbb}\,\bigg)\,,\label{gamma-k}
\eal
where $\omega=\esp{i 2\pi/3}$. Specifically, $\gamma(\bk)\sim A_1$ is invariant under $\Cg{3z}$, while 
\be
\gamma_{\pm 1}\Big(\Cg{3z}(\bk)\Big)=\omega^{\pm 1}\,\gamma_{\pm 1}(\bk)\,.
\ee
In other words, $\big(\gamma_{+1}(\bk),\gamma_{-1}(\bk)\big)$ 
form a representation of the 2d-irrep $E=\big(E_{+1},E_{-1}\big)$ in which $\Cg{3z}$ is diagonal
with eigenvalues $\omega$ and $\omega^*$.

\begin{figure}
\centerline{\includegraphics[width=0.35\textwidth]{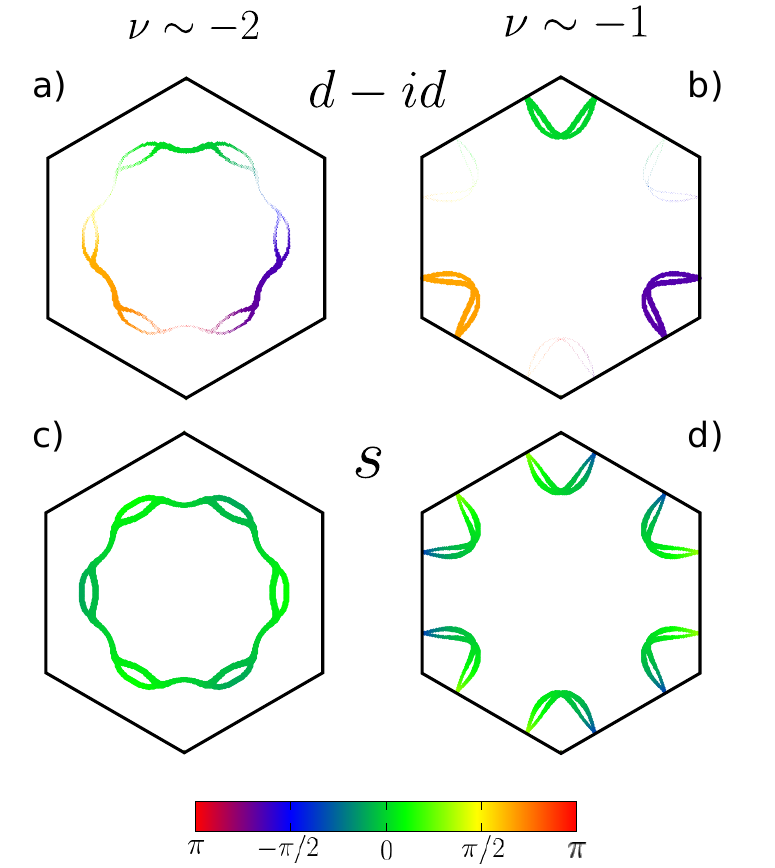}}
\caption{Pair amplitudes $\Delta(k)^\dagger$ of the leading  superconducting channels: $d-id$-wave (a-b) and extended $s$-wave (c-d).
In each panel we show the hexagonal MBZ and the phase amplitudes restricted to a narrow region close to the Fermi surfaces corresponding to the occupancies $\nu \sim -2$ (a-c) and $\nu \sim -1$  (b-d). The phase of the superconducting order parameter is expressed in color. Note that the double line are due to the fact that two bands cross the chemical potential at different $\bk$-points.}
\label{BCS}
\end{figure}

Here it is more convenient to transform the spinor $\Phi_{\bk\sigma}$
\beal
\begin{pmatrix}
\Phi_{+1,\bk\sigma}\\
\Phi_{-1,\bk\sigma}
\end{pmatrix}
=\fract{1}{\sqrt{2}}\,
\begin{pmatrix}
1 & -i\\
1 & + i
\end{pmatrix}\,\begin{pmatrix}
\Phi_{s,\bk\sigma}\\
\Phi_{p,\bk\sigma}
\end{pmatrix} \,,
\eal
so that $\Phi^\dagga_{\pm 1,\bk\sigma}$ is associated with a WO that 
transforms like $E_{\pm 1}$. Under the assumption of pairing diagonal in the 
irreps, we can construct the following spin-singlet Cooper pairs:
\bea
\sum_\sigma\,\sigma\,\Phi^\dagger_{+1,AB,\bk\sigma}\,\tau_1\,\Phi^\dagger_{+1,BA,-\bk-\sigma}
&\sim& E_{-1,\bk}\,,\label{E-1-SC}\\
\sum_\sigma\,\sigma\,\Phi^\dagger_{-1,AB,\bk\sigma}\,\tau_1\,\Phi^\dagger_{-1,BA,-\bk-\sigma}
&\sim & E_{+1,\bk}\,,\label{E+1-SC}\\
\fract{1}{\sqrt{2}}\sum_\sigma \sigma\Big(
\Phi^\dagger_{+1,AB,\bk\sigma}\,\tau_1\,\Phi^\dagger_{-1,BA,-\bk-\sigma}
 &&\\
\qquad
+\!(-) \Phi^\dagger_{-1,AB,\bk\sigma}\,\tau_1\,\Phi^\dagger_{+1,BA,-\bk-\sigma}
\Big) &\sim& A_{1(2),\bk}\,,\qquad \label{E-A}\\
\fract{1}{\sqrt{2}}\sum_\sigma\,\sigma\,\Psi^\dagger_{AB,\bk\sigma}\,\tau_1\,\mu_{0(3)}\,\Psi^\dagger_{BA,-\bk-\sigma}
&\sim& A'_{1(2),\bk}\,,\qquad\label{A-A}
\eea 
which can be combined with the $\bk$-dependent functions in \eqn{gamma-k} 
to give pair operators that transform like the irreps of $D_3$. 
For instance, multiplying \eqn{E-1-SC} by $\gamma_{+1}(\bk)$, 
\eqn{E+1-SC} by $\gamma_{-1}(\bk)$, \eqn{E-A} with 
the plus sign by $\gamma(\bk)$, or  
\eqn{A-A} with $\mu_0$ by $\gamma(\bk)$, we obtain pair operators that all 
transform like $A_1$. We shall denote their sum as $\textbf{A}^\dagger_{1\bk}$, and, similarly, all other symmetry combinations as 
$\textbf{A}^\dagger_{2\bk}$ and $\textbf{E}^\dagger_{\pm 1,\bk}$. Evidently, 
since in our modeling the FBs are made of 1d and 2d irreps, the gap function will 
in general involve a combination of $\gamma(\bk)$ and $\gamma_{\pm 1}(\bk)$, namely, 
it will be a superposition of $s$ and $d\pm i d$ symmetry channels. The dominant channel 
will depend on which is the prevailing WO character of the Bloch states at the chemical potential, as well as on the strength of the scattering amplitudes in the different pairing channels, 
$\textbf{A}^\dagger_{1\bk}$, $\textbf{A}^\dagger_{2\bk}$ and 
$\textbf{E}^\dagger_{\pm 1,\bk}$.   \\
In general, we may expect the totally-symmetric $A_1$ channel to have the largest amplitude, thus we
assume the following expression of the phonon-mediated attraction 
\bea
H_\text{el-el} &\simeq& -\fract{\lambda}{V}\sum_{\bk\bp}
\textbf{A}^\dagger_{1\bk}\,\textbf{A}^\dagga_{1\bp}
\,,\label{el-el}
\eea 
that involves a single parameter $\lambda\sim g^2/2\omega$.
We treat the full Hamiltonian \eqn{Ham} plus \eqn{el-el} in mean field allowing for a superconducting solution, which is always stabilized by the attraction provided the density of states is finite at the chemical potential. 
We find that superconductivity opens a gap everywhere in the Brillouin zone. Since 
\beal
\langle  \,\textbf{A}^\dagger_{1\bk}\,\rangle &= 
\gamma_{+1}(\bk)\,\langle\,E_{-1,\bk}\,\rangle 
+ \gamma_{-1}(\bk)\,\langle\,E_{+1,\bk}\,\rangle \\
&\quad + \gamma(\bk)\,\langle\,A_{1,\bk}\,\rangle
+ \gamma(\bk)\,\langle\,A'_{1,\bk}\,\rangle\,,
\eal
the order parameter may have finite components with different symmetries, $E_{\pm 1}$ and $A_1$. In the model calculation all components acquire similar magnitude, implying 
a mixture of $s$ and $d\pm id$ wave symmetries. In Fig.~\ref{BCS} we show 
$\langle\,A'_{1,\bk}\,\rangle$ and $\langle\,E_{-1,\bk}\,\rangle$ at the Fermi surface corresponding to 
densities $\nu\approx-1$ and $\nu\approx-2$ with respect to charge neutrality.
We conclude by emphasizing that the Cooper pair is made by one electron in AB and one in 
BA, thus leading, in the spin-singlet channel, to extended $s$ and/or $d\pm id$ symmetries.
That is merely a consequence of the phonon mode and electron-phonon 
properties, hence it does not depend on the above modeling of the FBs. \\

There are already in literature several 
proposals about the superconducting states in tBLG. Most of them, however, invoke electron correlations as the element responsible, or strongly effecting the pairing  \cite{Balents,Senthil-2,Baskaran,Dodaro,Fu_SC_PRX,Vishvanath_SC_IVC,Kennes,Liu,Zhang_SC,Spaek,Laksono,Lin,Adam_IVC_SC,Stauber-1,Stauber-2,Roy,Classen,Guo,Ma,Nandikshore}. 
There are  few exceptions \cite{McDonald_phBCS,Bernevig_phBCS,Sarma} that instead propose, as we do, a purely phonon mediated attraction. In both works \cite{McDonald_phBCS} and \cite{Bernevig_phBCS} the tBLG phonons are assumed 
to coincide with the single layer graphene ones, as if the interlayer coupling were ineffective in the phonon spectrum, which is not what we find for the 
special modes discussed above. Moreover, they both discuss the effects of such phonons only in a continuum model for the FBs.  
In particular, the authors of \cite{McDonald_phBCS} consider a few selected graphene modes, among which the transverse optical mode at $\bK$ that has the largest weight in the tBLG phonon that we consider. They conclude that such graphene mode mediates $d\pm id$ pairing in  
the $A_2$ channel, $\tau_1\,\mu_3$ in our language, leading to an order parameter odd upon interchanging the two layers. On the contrary, the authors 
of \cite{Bernevig_phBCS} focus just on the acoustic phonons of graphene, and 
conclude they stabilize an extended $s$-wave order parameter.  

\newpage

\section{Conclusions}
\label{Conclusions}
In this work, we uncovered a novel and strong electron-phonon coupling mechanism and analysed 
its potential role in the low temperature physics of magic angle twisted bilayer graphene, with particular emphasis on the insulating 
and superconducting 
states.
By working out the phonon modes of a fully relaxed tBLG, we found a special group of few of them modulated over the whole moir\'e supercell and nearly dispersionless, thus showing that the moir\'e pattern can induce flat bands also in the phonon spectrum.
In particular, two of these modes, which are degenerate at any $\bk$-point invariant under 
$180^\circ$ rotation around the $y$-axis and its $\Cg{3z}$ equivalent directions, are found to couple strongly to the valley $U_v(1)$ symmetry, 
that is responsible for  the accidental degeneracy of the band structure 
at the same $\bk$-points where the two modes are degenerate. 
This particular phonon doublet is strongly 
Jahn-Teller coupled to the valley degrees of freedom, realizing a so-called $\text{E}\otimes\text{e}$ Jahn-Teller model.
This mechanism, if static, would generate a filling-dependent broadening of the flat bands and eventually insulating phases at all the commensurate fillings.
Interestingly, freezing the modes at 
$\bGamma$ can stabilize a topologically non-trivial insulator at charge neutrality that sustains edge modes. 
We also investigate the symmetry properties of a hypothetical superconducting state stabilized by this 
Jahn-Teller mode. We find that the phonon mediated coupling  occurs on the moir\'e scale, favoring spin-singlet pairing of electrons in different Bernal (AB/BA) regions, which may thus condense with an extended $s$- and/or $d\pm id$-wave order parameter. In a mean-field calculation with a model tight-binding Hamiltonian of the 
flat bands, we find that the dominant symmetry depends on the 
orbital character that prevails in the Bloch states at the Fermi energy, as well as on the precise values of scattering amplitudes in the different Cooper channels allowed by the $\Cg{3z}$ symmetry. We cannot exclude that a nematic component might arise due to higher order terms not included in our mean-field calculation, as discussed in \cite{Sigrist&Ueda} and \cite{Kozii-nematic}.
These results herald a role of phonons 
and of lattice distortions of much larger impact than supposed so far in twisted graphene bilayers, 
which does not exclude 
a joint action of electron-phonon and electron-electron interaction. 
Further experimental and theoretical developments will be called for in order to establish their actual importance and role.

\section*{Acknowledgments}

The authors are extremely grateful to D. Mandelli for the technical support provided during  the lattice relaxation procedure. We acknowledge useful discussions with P. Lucignano, A. Valli, A. Amaricci, M. Capone, E. Kucukbenli, A. Dal Corso and S. de Gironcoli. M.~F. acknowledges funding by the European Research Council (ERC) under H2020 Advanced Grant No. 692670 ``FIRSTORM''.  E.~T. acknowledges funding from the European Research Council (ERC) under FP7 Advanced Grant No.320796  ``MODPHYSFRICT'' ,
later continuing 
under Horizon 2020 Advanced Grant  No. 824402 ''ULTRADISS''.

\newpage

\appendix
\section{Details on the lattice relaxation, band structure and phonon calculations}
\label{AppendixA}
The lattice relaxation and bandstructure calculation procedures are the same as those  thoroughly described in Ref.~\cite{Fabrizio}, with the exception that now the carbon-carbon intralayer interactions are modeled via the Tersoff potential \cite{TERSOFF-2010}. 
The interlayer interactions are modelled 
via the Kolmogorov-Crespi (KC) potential \cite{Kolmogorov-PPRB2005},
using the recent parametrization of Ref.~\onlinecite{Ouyang-NanoLett2018}. 
Geometric optimizations are performed using the FIRE algorithm \cite{Bitzek-PRL2006}.
The hopping amplitudes of the tight-binding Hamiltonian are:
\beal
t(\mathbf{d})= V_{pp\sigma}(d)\bigg[\frac{\textbf{d}\cdot \textbf{e}_z}{d}\bigg]^2 \!\!\!+V_{pp\pi}(d)
\bigg[1-\Big(\frac{\textbf{d}\cdot \textbf{e}_z}{d}\Big)^2\bigg]\,,
\label{A1}
\eal
where $\mathbf{d}= \br_i-\br_j$ is the distance between atom $i$ and $j$, $d=|\mathbf{d}|$, and $\textbf{e}_z$ is the unit vector in the direction perpendicular to the graphene planes.
The out-of-plane ($\sigma$)  and in-plane ($\pi$) transfer integrals are:
\beal
V_{pp\sigma}(x)=V_{pp\sigma}^0 \esp{-\frac{x-d_0}{r_0}}\;\;\;\;V_{pp\pi}(x)
=V_{pp\pi}^0 \esp{-\frac{x-a_0}{r_0}}
\eal
 where $V_{pp\sigma}^0=0.48\;eV$ and $V_{pp\pi}^0=-2.8\; eV$ are values chosen to reproduce ab-initio dispersion curves in AA and AB stacked bilayer graphene, $d_0=3.344 \text{\AA}$ is the starting inter-layer distance, 
$a_0=1.3978 \text{\AA}$ is the intralayer carbon-carbon
distance obtained with the Tersoff potential, and $r_0=0.3187~a_0$ 
is the decay length \cite{Trambly,Nam_Koshino_PRB}.
\\

\begin{figure}
\centerline{\includegraphics[width=0.35\textwidth]{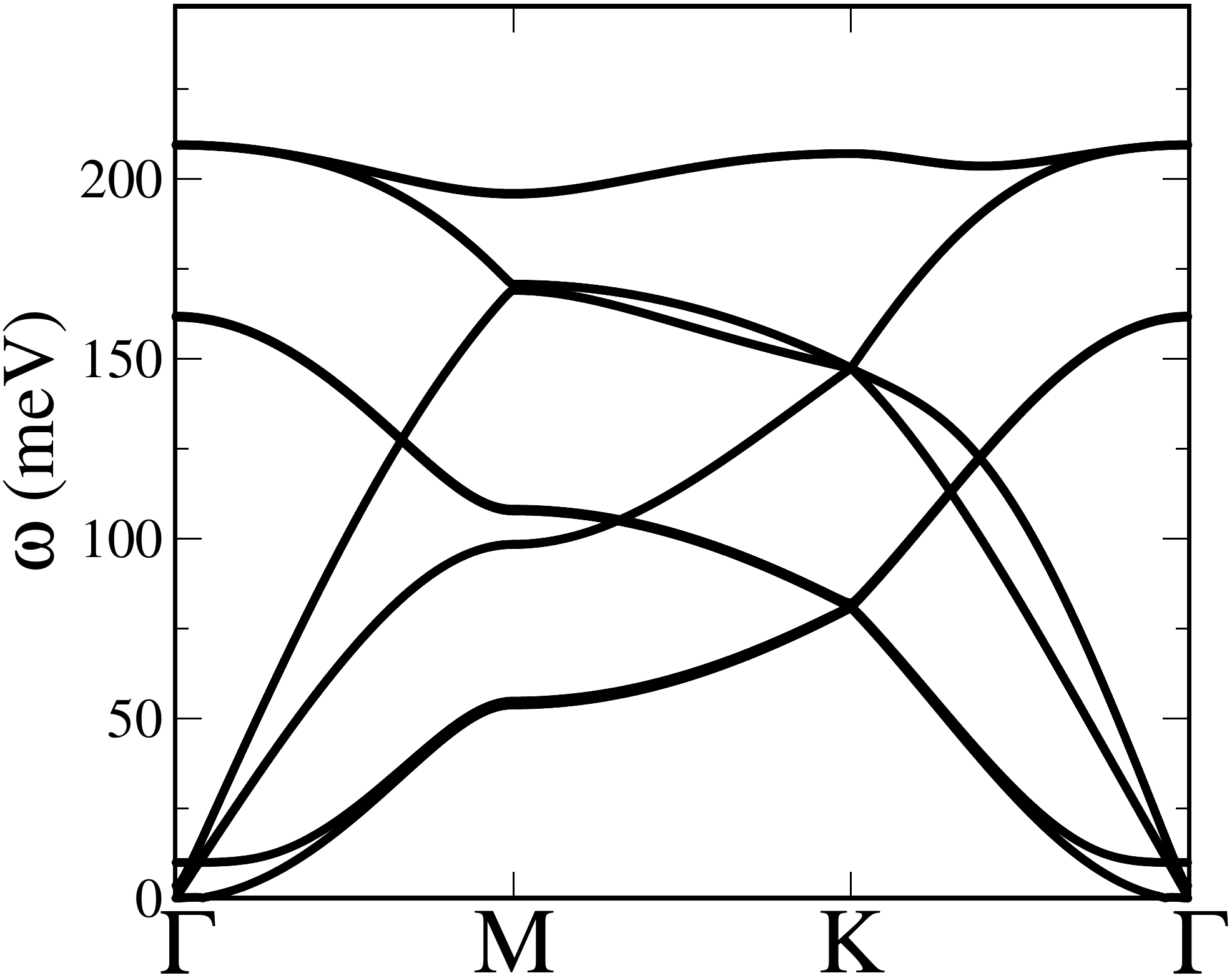}}
\caption{Phonon dispersion obtained with our choice of intralayer and interlayer potentials in Bernal stacked bilayer graphene. }
\label{ph_bernal}
\end{figure}

We compute phonons in tBLG using the force constants of the non-harmonic potentials $U=U_{TERSOFF}+U_{KC}$  that we used to relax the structure:
\begin{equation}
C_{\alpha \beta}(il,js)= \frac{\partial^2 U}{\partial R_{\alpha il}\partial R_{\beta js}}
\end{equation}
where ($i,j$) label the atoms in the unit cell, ($l,s$) the moir\'e lattice vectors and 
$\alpha,\beta=x,y,z$.
Then, we define the Dynamical Matrix at phonon momentum $\bq$ as:
\begin{equation}
D_{\alpha \beta i j}(\bq)=\frac{1}{M_C}\sum_{l} C_{\alpha\beta}(il,j0) e^{-i\bq R_l}
\end{equation}
where $M_C$ is the carbon atom mass.
Using the above relation we determine the eigenvalue equation for the normal modes of the system and the phonon spectrum:
\begin{equation}
\sum_{j\beta} D_{\alpha \beta i j}(\bq)\epsilon_{j\beta}(\bq)=\omega_{\bq}^2\epsilon_{i\alpha}(\bq)
\end{equation}
where $\omega_\bq$ denote the energy of the normal mode $\epsilon(\bq)$.
The phonon dispersion of Bernal stacked bilayer graphene obtained with this method is shown 
Fig.~\ref{ph_bernal}. 
As can be seen, the transverse optical (TO) modes at $K$ are found at $\omega\approx 207$ meV. As a consequence, the Jahn-Teller modes discussed in the main text, which vibrate in the same way on the graphene scale, have similar frequency. However, the frequency of the TO modes in graphene is strongly sensitive to the choice of the intralayer potential used \cite{MD-graphene-phonons}, so that these modes can be predicted to have frequencies as low as $\approx 170$ meV \cite{Nika-phonon,Basko-phonon}. This implies that also the JT modes may be observed at a lower frequencies than ours. 

\section{Hamiltonian in momentum space}
\label{AppendixB}
In order to make the invariance under $\Cg{3z}$ more explicit, 
we shall use the transformed spinors 
\beal
\begin{pmatrix}
\Phi_{+1,\bk\sigma}\\
\Phi_{-1,\bk\sigma}
\end{pmatrix}
=\fract{1}{\sqrt{2}}\,
\begin{pmatrix}
1 & -i\\
1 & + i
\end{pmatrix}\,\begin{pmatrix}
\Phi_{s,\bk\sigma}\\
\Phi_{p,\bk\sigma}
\end{pmatrix} \,,
\eal
for the 2d-irreps, which correspond to WOs eigenstates of 
$\Cg{3z}$. Moreover, it is convenient to transform also the 
spinors $\Psi_{\bk\sigma}$ of the 1d-irreps in the same way, i.e., 
\beal
\begin{pmatrix}
\Psi_{+1,\bk\sigma}\\
\Psi_{-1,\bk\sigma}
\end{pmatrix}
=\fract{1}{\sqrt{2}}\,
\begin{pmatrix}
1 & -i\\
1 & + i
\end{pmatrix}\,\begin{pmatrix}
\Psi_{s,\bk\sigma}\\
\Psi_{p,\bk\sigma}
\end{pmatrix} \,,
\eal
which correspond to WOs still invariant under $\Cg{3z}$ but not under 
$\Cg{2x}$, whose representation both in $\Psi_{\bk\sigma}$ and 
$\Phi_{\bk\sigma}$ becomes the Pauli matrix $\mu_1$. In conclusion the spinor 
operators defined above satisfy  
\beal
\Cg{3z}\Big(\Phi_{\bk\sigma}\Big)
&= \begin{pmatrix}
\omega & 0\\
0 & \omega^*
\end{pmatrix}\,\Phi_{\Cg{3z}(\bk)\,\sigma}\,,\\
\Cg{3z}\Big(\Psi_{\bk\sigma}\Big)
&= \Psi_{\Cg{3z}(\bk)\,\sigma}\,,\\
\Cg{2x}\Big(\Phi_{\bk\sigma}\Big)
&= \mu_1\,\Phi_{\Cg{3z}(\bk)\,\sigma}\,,\\
\Cg{2x}\Big(\Psi_{\bk\sigma}\Big)
&= \mu_1\,\Psi_{\Cg{3z}(\bk)\,\sigma}\,.\label{App:field-transf}
\eal
\\
We shall, for simplicity, consider only nearest and next nearest neighbor 
hopping between AB and BA region, which correspond to the 
following functions in momentum space:
\beal
\gamma_1(\bk) &= \alpha_\bk\,
\bigg(1+ \esp{-i\bk\cdot\bba} + \esp{-i\bk\cdot\bbb}\,\bigg)\,,\\
\gamma_{1,+1}(\bk) &= \alpha_\bk\,
\bigg(1+ \omega\,\esp{-i\bk\cdot\bba} + \omega^*\,\esp{-i\bk\cdot\bbb}\,\bigg)\,,\\
\gamma_{1,-1}(\bk) &= \alpha_\bk\,
\bigg(1+ \omega^*\,\esp{-i\bk\cdot\bba} + \omega\,\esp{-i\bk\cdot\bbb}\,\bigg)\,,\label{App:gamma-k}
\eal
for first neighbors, and 
\beal
\gamma_2(\bk)\! &=\alpha_\bk\bigg(
\esp{i\bk\cdot(\bba-\bbb)} + \esp{-i\bk\cdot(\bba-\bbb)} +\esp{-i\bk\cdot(\bba+\bbb)} \bigg)\,,\\
\gamma_{2,+1}(\bk)\! &=\alpha_\bk\bigg(\!
\omega\,\esp{i\bk\cdot(\bba-\bbb)}\! + \omega^*\esp{-i\bk\cdot(\bba-\bbb)} 
\!+\esp{-i\bk\cdot(\bba+\bbb)} \bigg)\,,\\
\gamma_{2,-1}(\bk)\! &=\alpha_\bk\bigg(\!
\omega^*\esp{i\bk\cdot(\bba-\bbb)}\! + \omega\,\esp{-i\bk\cdot(\bba-\bbb)} 
\!+\esp{-i\bk\cdot(\bba+\bbb)} \bigg)\,,
\eal  
for second neighbors,
where $\omega=\esp{i 2\pi/3}$, $\alpha_\bk = \esp{i\bk\cdot(\bba+\bbb)/3}$, 
and the lattice constants $\bba=(\sqrt{3}/2,-1/2)$ and $\bbb=(\sqrt{3}/2,1/2)$.
Since 
\beal
\Cg{3z}\big(\bba\big) &= \bbb-\bba\,,& 
\Cg{3z}\big(\bbb\big) &= -\bba\,,\\
\Cg{2x}\big(\bba\big) &= \bbb\,,& 
\Cg{2x}\big(\bbb\big) &= \bba\,,
\eal
then, for $n=1,2$, 
\beal
\gamma_{n}\Big(\Cg{3z}(\bk)\Big) &= \gamma_{n}(\bk)\,,\\
\gamma_{n,\pm 1}\Big(\Cg{3z}(\bk)\Big) &= \omega^{\pm 1}\,\gamma_{n,\pm 1}(\bk)\,,\\
\gamma_{n}\Big(\Cg{2x}(\bk)\Big) &= \gamma_{n}(\bk)\,,\\
\gamma_{n,\pm 1}\Big(\Cg{2x}(\bk)\Big) &= \gamma_{n,\mp 1}(\bk)
\,,\label{App:gamma-transf}
\eal
which shows that $\gamma_{n,\pm 1}(\bk)$ transform like the 2d-irrep $E$.\\

We assume the following tight-binding Hamiltonian for the 1d-irreps     
\bea
H_{1d-1d} &=& \sum_{\bk\sigma}\bigg[
-\Delta\,\Psi^\dagger_{\bk\sigma}\,\sigma_0\,\mu_1\,\tau_0\,
\Psi^\dagga_{\bk\sigma}\label{App:1d-1d}\\
&& -\!\sum_{n=1,2}\!t^{(n)}_{11}\,
\Big(\gamma_n(\bk)\,\Psi^\dagger_{\bk\sigma}\,
\sigma^+\,\mu_0\,\tau_0\,\Psi^\dagga_{\bk\sigma} + H.c.\Big)
\bigg],\nonumber
\eea
where $t^{(1)}_{11}$ and $t^{(2)}_{11}$ are the first and second neighbor 
hopping amplitudes, respectively, which we assume to be real. \\
The 2d-irreps have instead the Hamiltonian 
\bea
H_{2d-2d} &=& -\sum_{\bk\sigma}\,\sum_{n=1}^2\,\Bigg[
t^{(n)}_{22}\,\gamma_n(\bk)\,\Phi^\dagger_{\bk\sigma}\,
\sigma^+\,\mu_0\,\tau_0\,\Phi^\dagga_{\bk\sigma} \nonumber\\
&&\! + g^{(n)}_{22}\,\Phi^\dagger_{\bk\sigma}\,
\sigma^+\,\hat\gamma_n(\bk)\,\mu_1\,\tau_0\,\Phi^\dagga_{\bk\sigma} + H.c.
\Bigg],\qquad\label{App:2d-2d}
\eea  
with real hopping amplitudes, where 
\be
\hat\gamma_n(\bk)
= \begin{pmatrix}
\gamma_{n,+1}(\bk) & 0\\
0 & \gamma_{n,-1}(\bk)
\end{pmatrix}\,.
\ee
Finally, the coupling between 1d and 2d irreps is represented by 
the Hamiltonian 
\bea
H_{1d-2d} &=& -\sum_{\bk\sigma}\,\sum_{n=1}^2\,
t^{(n)}_{12}\Bigg[
\Phi^\dagger_{\bk\sigma}\,\sigma^+\,\mu_1\,\hat\gamma_n(\bk)\,
\mu_1\,
\tau_0\,\Psi^\dagga_{\bk\sigma}\nonumber\\
&& + \Psi^\dagger_{\bk\sigma}\,\sigma^+\,\hat\gamma_n(\bk)\,
\tau_0\,\Phi^\dagga_{\bk\sigma}\nonumber\\
&& + i\,\Phi^\dagger_{\bk\sigma}\,\sigma^+\,\mu_1\,\hat\gamma_n(\bk)\,
\tau_3\,\Psi^\dagga_{\bk\sigma}\label{app:1d-2d}\\
&&+ i\,\Psi^\dagger_{\bk\sigma}\,\sigma^+\,\mu_1\,\hat\gamma_n(\bk)\,
\tau_3\,\Phi^\dagga_{\bk\sigma} + H.c.\Bigg]\nonumber\,,
\eea
 with real $t^{(n)}_{12}$. \\
 The Hamiltonian thus reads 
 \beal
 H &= H_{1d-1d}+H_{2d-2d}+H_{1d-2d}\,,\label{App:Ham}
 \eal
which, through the equations 
 \eqn{App:field-transf} and \eqn{App:gamma-transf}, can be readily shown to be 
invariant under $\Cg{3z}$ and $\Cg{2x}$, and is evidently also invariant under the 
$U_v(1)$ generator $\tau_3$. In addition, the Hamiltonian must be also invariant 
under $\text{T}\Cg{2z}$, where T is the time reversal operator. Noting that
\beal
\text{T}\Cg{2z}\Big(\Phi_{\bk\sigma}\Big) &= 
\sigma_1\,\mu_1\,\Phi_{\bk-\sigma}\,,\\
\text{T}\Cg{2z}\Big(\Psi_{\bk\sigma}\Big) &= 
\sigma_1\,\mu_1\,\Psi_{\bk-\sigma}\,,
\eal
one can show that $H$ in \eqn{App:Ham} is also invariant under that symmetry. 
 \\

The model Hamiltonian thus depends on eight parameters. The FBs shown in 
Fig.~\ref{sketch} have been obtained choosing: 
$\Delta=10$ , $t_{11}^1=2$ , $t_{22}^1=5$ , $g_{22}^1=10$ , $t_{22}^{2}=g_{22}^2=-t_{11}^2=1.2$  , $t_{12}^1=2$ and $t_{12}^2=0.5$.

\bibliographystyle{apsrev}

\end{document}